\documentclass[aps,prl,twocolumn,superscriptaddress]{revtex4-1}
\usepackage[utf8]{inputenc}
\usepackage{chemformula}
\usepackage{amsmath}
\usepackage{mathrsfs}
\usepackage{physics}
\usepackage{multirow}
\usepackage{amssymb}
\usepackage{graphicx}
\usepackage{siunitx}
\usepackage{lipsum}
\usepackage[normalem]{ulem}
\newcommand{\Tc}{$T_\mathrm{c}$ }
\newcommand{\ohm}{$\Omega$}

\definecolor{mag}{RGB}{255,0,255}
\definecolor{violet}{RGB}{148,0,211}

\begin{document}
\title{Creating superconductivity in \ch{WB2} through pressure-induced\\ metastable planar defects}

\author{J.\ Lim}
\thanks{These two authors contributed equally to this work.}
\affiliation{Department of Physics, University of Florida, Gainesville, Florida 32611, USA}
\author{A.\ C.\ Hire}
\thanks{These two authors contributed equally to this work.}
\affiliation{Department of Materials Science and  Engineering, University of Florida, Gainesville, Florida 32611, USA}
\affiliation{Quantum Theory Project, University of Florida, Gainesville, Florida 32611, USA}
\author{Y.\ Quan}
\affiliation{Department of Physics, University of Florida, Gainesville, Florida 32611, USA}
\affiliation{Department of Materials Science and  Engineering, University of Florida, Gainesville, Florida 32611, USA}
\affiliation{Quantum Theory Project, University of Florida, Gainesville, Florida 32611, USA}
\author{J.\ S.\ Kim}
\affiliation{Department of Physics, University of Florida, Gainesville, Florida 32611, USA}
\author{S.\ R.\ Xie}
\affiliation{Department of Materials Science and  Engineering, University of Florida, Gainesville, Florida 32611, USA}
\affiliation{Quantum Theory Project, University of Florida, Gainesville, Florida 32611, USA}
\author{S.\ Sinha}
\affiliation{Department of Physics, University of Florida, Gainesville, Florida 32611, USA}
\author{R.\ S.\ Kumar}
\affiliation{Department of Physcis, Chemistry, and Earth and Environmental Sciences, University of Illinois Chicago, Chicago, Illinois 60607, USA}
\author{D.\ Popov}
\affiliation{HPCAT, X-ray Science Division, Argonne National Laboratory, Argonne, Illinois 60439, USA}
\author{C.\ Park}
\affiliation{HPCAT, X-ray Science Division, Argonne National Laboratory, Argonne, Illinois 60439, USA}
\author{R.~J.~Hemley}
\affiliation{Department of Physcis, Chemistry, and Earth and Environmental Sciences, University of Illinois Chicago, Chicago, Illinois 60607, USA}
\author{Y.\ K.\ Vohra}
\affiliation{Department of Physics, University of Alabama at Birmingham, Birmingham, Alabama 35294, USA}
\author{J.\ J.\ Hamlin}
\email{jhamlin@ufl.edu}
\affiliation{Department of Physics, University of Florida, Gainesville, Florida 32611, USA}
\author{R.\ G.\ Hennig}
\affiliation{Department of Materials Science and  Engineering, University of Florida, Gainesville, Florida 32611, USA}
\affiliation{Quantum Theory Project, University of Florida, Gainesville, Florida 32611, USA}
\author{P.\ J.\ Hirschfeld}
\affiliation{Department of Physics, University of Florida, Gainesville, Florida 32611, USA}
\author{G.\ R.\ Stewart}
\affiliation{Department of Physics, University of Florida, Gainesville, Florida 32611, USA}

\date{\today}

\begin{abstract}
    High-pressure electrical resistivity measurements reveal that the mechanical deformation of ultra-hard \ch{WB2} during compression induces superconductivity above 50~GPa with a maximum superconducting critical temperature, \Tc of 17~K at 91~GPa.
    Upon further compression up to 187~GPa, the \Tc gradually decreases.
    Theoretical calculations show that electron-phonon mediated superconductivity originates from the formation of metastable stacking faults and twin boundaries that exhibit a local structure resembling \ch{MgB2} (hP3, space group 191, prototype AlB$_2$).
    Synchrotron x-ray diffraction measurements up to \SI{145}{GPa} show that the ambient pressure hP12 structure (space group 194, prototype \ch{WB2}) continues to persist to this pressure, consistent with the formation of the planar defects above 50~GPa.
    The abrupt appearance of superconductivity under pressure does not coincide with a structural transition but instead with the formation and percolation of mechanically-induced stacking faults and twin boundaries. The results identify an alternate route for designing superconducting materials.
\end{abstract}

\maketitle

\section{Introduction}
In 2001 superconductivity with the remarkably high critical temperature of \SI{39}{K} was discovered in \ch{MgB2}.
Efforts to increase \Tc beyond the ambient pressure value in the material invariably proved unsuccessful, as both pressure~\cite{tomita_dependence_2001,deemyad_dependence_2003} and various chemical substitutions~\cite{buzea_review_2001,budko_superconductivity_2015} caused a decrease in the critical temperature.
After two decades of searching for further high \Tc superconductors in the diboride family of compounds, Pei {\it et al.}~\cite{MoB2_superconductivity} recently discovered that \ch{MoB2} transforms from the hR6, ($R\bar 3m$, CaSi${_2}$) structure to the hP3 ($P6/mmm$, \ch{AlB2} or \ch{MgB2}) structure above \SI{70}{GPa}, and exhibits a \Tc that reaches \SI{32}{K} at \SI{100}{GPa}.

Inspired by this result, we have synthesized single crystals of the isoelectronic compound \ch{WB2}, which forms at ambient pressure in the hP12 ($P6_3/mmc$) structure. Investigating this compound in a series of experiments to pressures as high as \SI{187}{GPa}, we discovered that \ch{WB2} becomes superconducting near \SI{50}{GPa}, with \Tc jumping  rapidly to about \SI{17}{K} and varying weakly with pressure thereafter.

Interestingly, \ch{WB2} had been studied earlier at ambient pressure, with results that differ from ours.
A critical temperature of \Tc = \SI{5.4}{K} was reported in Ref.~\onlinecite{Kahyan2012}, along with extensive x-ray and neutron diffraction data, leading to a suggestion that the hP12 structure was realized in  large grain polycrystalline samples.
This work emphasized  that \ch{WB2} had previously been identified as \ch{W2B5}~\cite{Frotscher2007}.
The reason for the difficulty of extracting the correct structure in diffraction patterns is the enormous $Z$-contrast between W (74) and B (5), such that the B positions are difficult to ascertain.
Therefore, theoretical studies are needed to resolve the origin of the reported ambient pressure superconductivity, the difference with our results, and the jump of $T_c$ to \SI{17}{K} at \SI{91}{GPa}.
By themselves, our resistivity data might suggest a transition to a new structure. However, synchrotron x-ray diffraction measurements under pressure indicate that the material, or a major part of it, retains the ambient pressure hP12 structure with a monotonically increasing $c/a$ ratio.
Given the weak scattering of x-rays by B, small contributions from hR6 and hP3 may be present.

To explain the experimental observations, we explored the relative stability of relevant phases and investigated their superconducting properties as a function of pressure using density-functional theory (DFT).
The ambient pressure hP12 phase is found to have nearly the lowest enthalpy up to pressures of 120 GPa, whereas the \ch{MgB2}-like hP3 phase is strongly disfavored over most of this range.
However, the hP12 phase was calculated to have a very low critical temperature (\Tc $<1$~K) over these pressures.
On the other hand, the metastable hP3 phase is predicted to have a critical temperature of 25-40~K over a wide pressure range.
Thus, while the hP3 structure might explain the finding of higher $T_c$, it is never sufficiently stable.

To resolve this paradox, we have investigated defect structures intermediate between the hP3 and hP12 phases, i.e., those involving twin boundaries and stacking faults.
Based on their formation enthalpies, we estimate that such planar defects plausibly occur within the hP12 phase during plastic deformation of the sample.
These defect structures resemble nanometer-thick regions of the hP3 and hR6 structures, with W atoms in eclipsing positions above and below unbuckled B hexagons.
We thus argue that the observed superconductivity appears following the formation of significant quantities of stacking faults above 50 GPa, which percolate through the sample above 100 GPa.
In contrast to our results obtained on single crystals, the presence of hR6-like planar defects may also explain the superconductivity observed at ambient pressure in large-grain polycrystalline samples~\cite{Kahyan2012}.

\begin{figure*}[t!]
    \centering
    \includegraphics[width=0.75\textwidth]{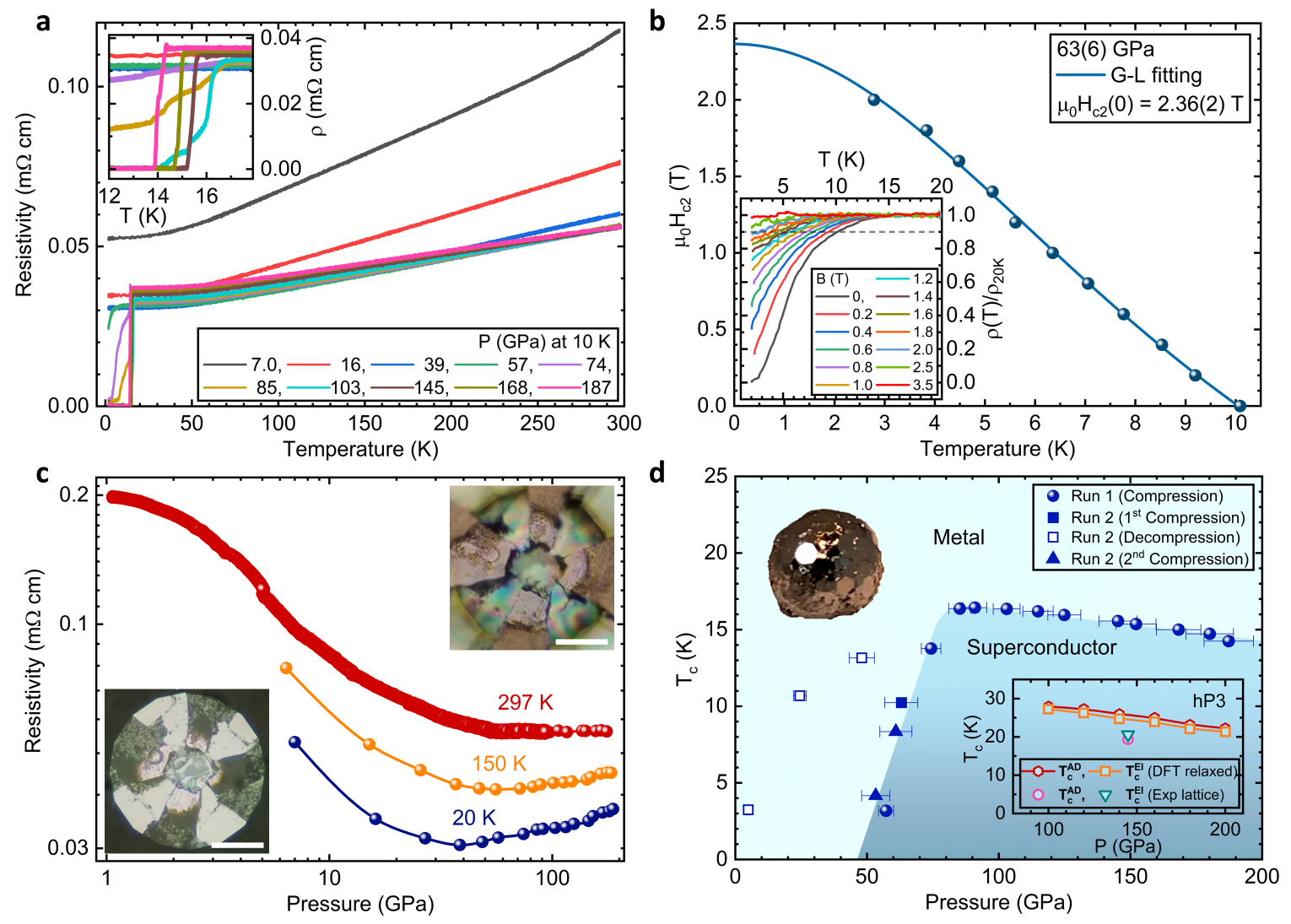}
    \caption{Representative high-pressure resistivity curves and $T_{c}$(P) phase diagram. \textbf{a} Temperature-dependent resistivity curves of \ch{WB2} measured while warming at several pressures to \SI{187}{GPa} during compression (7.0 and \SI{16}{GPa} data measured while cooling). The reported pressures are measured at \SI{10}{K}. The black crossed lines represent the criterion used to determine $T_{c}$ (onset). The inset shows an enlarged view of the superconducting transition. \textbf{b} Temperature-dependent upper critical field of \ch{WB2} at \SI{63}{GPa} (Run 2). The dark cyan solid line refers to Ginzburg-Landau (G-L) fitting. The inset shows temperature-dependent relative resistivity curves (after smoothing) under different magnetic fields to \SI{3.5}{T}. The gray dashed line indicates 90\% of the normal state value, where the $T_{c}$ is defined for the upper critical field. \textbf{c} Pressure-dependent resistivity curves at room temperature (RT), 150, and \SI{20}{K}. The insets show photographs of the experimental setup at $\sim$1 (lower left) and \SI{173}{GPa} (upper right), respectively. The white scale bars indicate \SI{50}{\micro\meter}. \textbf{d} Superconducting phase diagram of \ch{WB2} with pressure (at \SI{10}{K}) using the 90\% criterion. Closed dark blue symbols (spheres, a square, and triangles) indicate the $T_c$ taken from compression (Run 1 and Run 2), whereas open blue square symbols are from decompression (Run 2). A sharp increase of $T_{c}$ is present around \SI{70}{GPa}. The inset in the upper left shows a photograph of an arc melted boule of \ch{WB2}. Hexagonal crystal facets are visible on the surface. A small single crystalline fragment of the larger boule was selected for the electrical resistivity measurements under pressure. A second inset at lower right shows the Eliashberg theory $T_c$ calculated for WB$_2$ in the hP3 structure with DFT relaxed lattice constants at 100, 120, 140, 160 and 180 GPa (orange squares), along with the Allen-Dynes $T_c$ (red hexagons). A magenta circle and a green upside-down triangle show similar calculations for experimental lattice constants at 145 GPa.}
    \label{fig:WB2_rhoPT}
\end{figure*}

This unprecedented creation of superconductivity through mechanically induced metastable defects opens opportunities to search for other materials systems where metastable structures can be stabilized in the form of planar defects.
While interfaces and twin boundaries can lead to surface phonons that somewhat increase an existing $T_c$~\cite{Buzdin1988}, the mechanism here involves the formation of percolating metastable planar defects induced by mechanical processing, thereby inducing superconductivity.
This offers the potential to stabilize at ambient pressure defect microstructures that exhibit novel properties.
With the discovery of high-$T_c$ superconductivity in high-pressure hydrides~\cite{h3sexp,Somayazulu}, new routes are being sought to stabilize metastable high-$T_c$ superconductivity~\cite{Boeri2021}, and our proposal may represent a promising direction towards this goal.

\begin{figure*}
    \centering
    \includegraphics[width=0.8\textwidth]{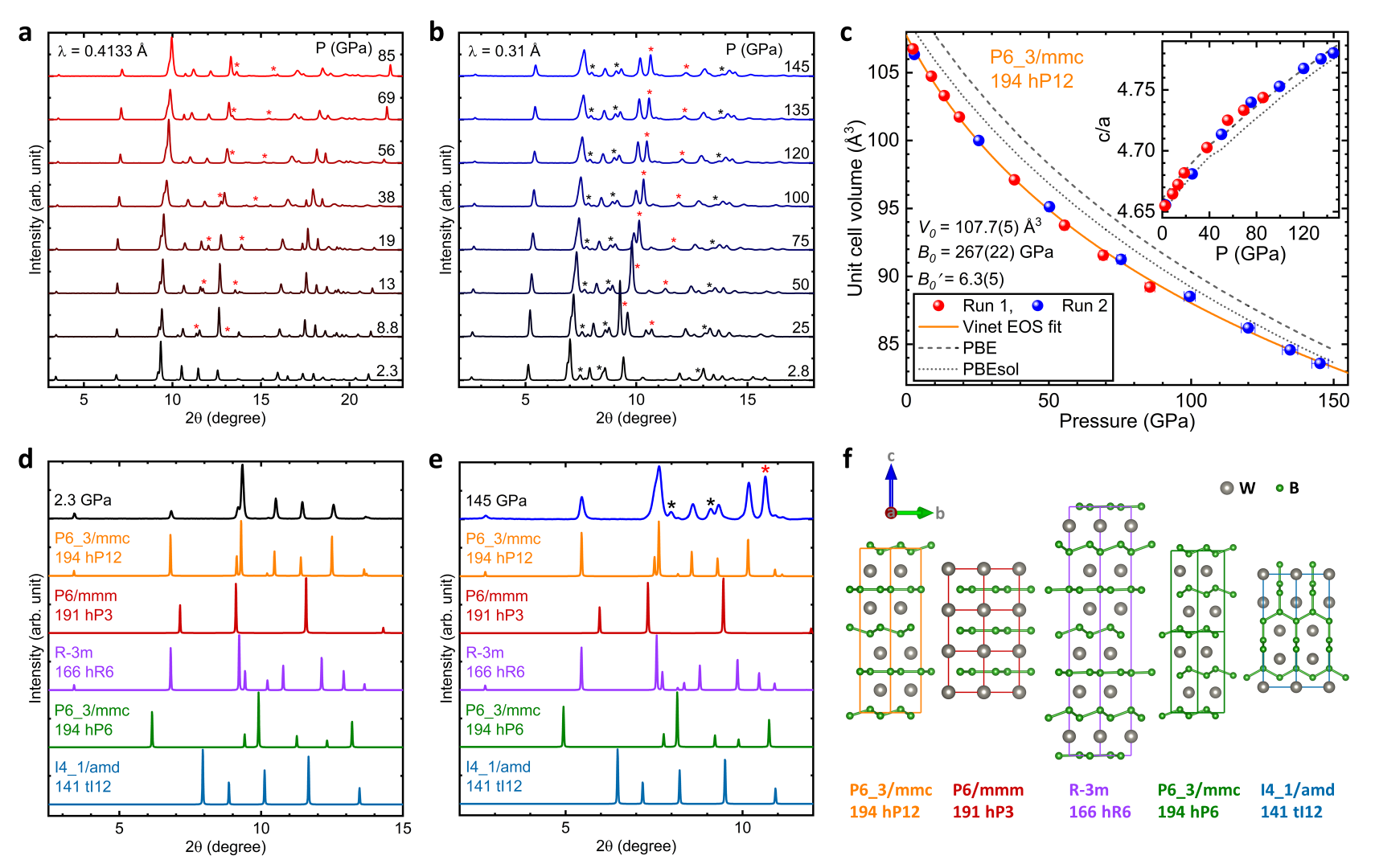}
    \caption{High-pressure XRD patterns, PV-isotherm, and crystal structure comparison. \textbf{a, b} High-pressure XRD patterns of \ch{WB2} obtained from run 1 and 2, respectively, at room temperature using \ch{Ne} as the quasi-hydrostatic pressure medium. No structural transition was observed throughout the pressure range studied. The star symbols denote the Bragg peaks of the Re gasket (black color) and the Ne pressure-transmitting medium (red color), respectively. \textbf{c} The resulting pressure-volume (P-V) curve fitted with the Vinet equation of state (EOS)~\cite{Vinet1987}. The dashed and dotted lines refer to theoretical calculations from PBE and PBEsol, respectively~\cite{PBE}. The inset shows the $c/a$ ratio versus pressure. 
    \textbf{d, e} Comparison of experimental XRD patterns and those of the theoretical structure models from Fig.~\ref{WB2_enthalpy} at 2.3 and 145 GPa. \textbf{f} Comparison of crystal structures showing different sequences along the $c$-axis. To aid comparison, the colors of the curves in \textbf{d} and \textbf{e} match those in the corresponding structure diagrams shown in \textbf{f}.
    \label{WB2_HPXRD}}
\end{figure*}

\section{Results}
\subsection{Experiment}
We synthesized high-quality single crystals of \ch{WB2} using an arc-melting technique described in the Methods section.
Figure~\ref{fig:WB2_rhoPT} presents a summary of the electrical resistivity measurements on  one of these samples,  extending from \SI{1.8}{K} to \SI{297}{K} and to pressures as high as \SI{187}{GPa}. The pressure values shown are measured at \SI{10}{K}. The residual-resistivity ratio (RRR) at 6 GPa is 2.24 calculated from $\rho$(\SI{297}{K}) = 0.118 m\ohm-cm over $\rho$(\SI{1.8}{K}) = 0.053 m\ohm-cm.
Superconductivity first appears at \SI{57}{GPa} as a broad, incomplete drop in the resistivity, with an onset near \SI{4}{K} (Fig.~\ref{fig:WB2_rhoPT}a) marked by two crossing lines for \Tc(onset).
With further pressure increase, \Tc goes up rapidly at a rate of \SI{0.72}{K/GPa}.
At \SI{74}{GPa} the transition onset has reached \SI{17}{K}, but the transition remains broad, with the resistance failing to reach zero at the lowest temperatures.
Additional pressure increases have only a weak effect on the superconducting transition onset temperature with a rate of \SI{-0.024}{K/GPa}, but the transition becomes much more sharp.
Zero resistance is achieved for pressures above 80 GPa.
The superconductivity of \ch{WB2} is further supported by the suppression of \Tc with increasing external magnetic fields as shown in Fig.~\ref{fig:WB2_rhoPT}b.
The temperature-dependent relative resistivity curves at \SI{63}{GPa} in Fig.~\ref{fig:WB2_rhoPT}b inset show that the superconducting transition is suppressed by increasing field and completely destroyed above \SI{3.5}{T}, where \Tc is defined as the temperature at which the resistance has dropped to to 90\% of the normal-state resistivity just above the transition.
The temperature-dependent upper critical fields are fitted using the empirical Ginzburg-Landau (G-L) formula, $\mu_0H_{c2}(T) = \mu_0H_{c2}(0)(1-t^{2})/(1+t^{2})$, where $t$ is the temperature normalized  to the zero-field superconducting transition temperature $T/T_{c0}$ and $\mu_0H_{c2}(0)$ is the zero-temperature upper critical field.
The G-L fitting provides \SI{2.36}{T} for $\mu_0H_{c2}(0)$ at \SI{63}{GPa}.
Figure~\ref{fig:WB2_rhoPT}c shows the electrical resistivity as a function of pressure at three different temperatures.
The room temperature resisitivity drops monotonically, but at the lowest temperatures (\SI{20}{K}) the resistivity, which is related to the residual resistivity due to the defect scattering, exhibits a broad minimum as a function of pressure.
This minimum appears to roughly coincide with the pressure at which superconductivity first appears above \SI{50}{GPa}.

The pressure dependence of $T_c$ is shown in Figure~\ref{fig:WB2_rhoPT}d using the 90\% criterion.
Below \SI{50}{GPa}, no trace of superconductivity is observed down to \SI{1.8}{K}.
Superconductivity suddenly appears above \SI{50}{GPa}, with \Tc rapidly rising to \SI{14}{K}.
On subsequent pressure increase, the transition temperature passes through a broad maximum with a maximum of \SI{17}{K} (onset) near \SI{90}{GPa} and then gradually declines.
The broad superconducting transitions in the pressure range between 50 and \SI{90}{GPa}, together with the sudden increase in the onset temperature in this range suggests that the transition temperature in fact increases discontinuously, and that the broad resistive transitions may be caused by incomplete percolation of the superconducting phase through the sample.
We note that broad, multi-step transitions have been observed in other systems in the vicinity of transitions between different structural/electronic phases~\cite{Zhang2017}.
The pressure-dependent superconducting transition temperature is shown be reproducible in Run 2 (see also Supplementary Figure 7) under compression to \SI{63}{GPa}.
Interestingly, under decompression to \SI{48}{GPa}, \Tc first \textit{increases} from $\sim$ \SI{10}{K} to \SI{13}{K} and then decreases to  with further decompression to \SI{4.9}{GPa}. This irreversible behavior suggests the superconducting phase of \ch{WB2} is metastable at low pressure. 
The subsequent second compression on the same sample in Run 2 after completely opening the cell and released to ambient pressure, the superconductivity reappears only above \SI{50}{GPa} (See Supplementary Figure 8 and~\ref{fig:WB2_rhoPT}d) consistent with the compression behavior in Run 1. These additional results suggest that the planar defects and twin boundaries responsible for superconductivity are only (meta)stable under high pressure.
The abrupt appearance of superconductivity with a large $T_c (P)$ slope, followed by a sudden change to a lower slope at higher pressures is suggestive of a possible structural transition.
However, the normal state electrical resistivity does not exhibit any clear features that could be attributed to a structural transition.
We carried out high-pressure x-ray diffraction measurements in order to determine if a structural transition could be responsible for the appearance of superconductivity.

Figure~\ref{WB2_HPXRD} summarizes the x-ray diffraction data that we have obtained on \ch{WB2} to pressures as high as 145 GPa using \ch{Ne} as the quasi-hydrostatic pressure medium.
The diffraction patterns in Fig.~\ref{WB2_HPXRD}a (run 1) and b (run 2) used different x-ray wavelengths (0.41 and \SI{0.31}{\angstrom}, respectively), and present a consistent result.
The patterns are well described by the ambient pressure hP12 structure ($P6_3/mmc$, 194)~\cite{Wang_WB2ambient_2017}.
Fig.~\ref{WB2_HPXRD}c shows the experimental unit-cell volume along with the theoretically obtained results relaxed using PBEsol and PBE functional for the DFT calculations.
The theoretically calculated volumes agree well with the experimental values, with a slight overestimation in the DFT calculated volumes, as  expected~\cite{PhysRevB.82.014101}.
The obtained bulk modulus ($B_0$) \SI{267}{GPa} from a Vinet equation of state fit is relatively low compared to a previous experimental study finding \SI{349}{GPa} from ambient pressure ultrasonic measurements~\cite{Yin_moduliWB2_2013}.

Figures~\ref{WB2_HPXRD}d and e show the comparison of experimental XRD patterns for \ch{WB2} at 2.3 and \SI{145}{GPa}, respectively, with the five different structure models calculated and shown in Fig.~\ref{WB2_enthalpy}.
The hP12 structure, which is found to be the ambient structure~\cite{Wang_WB2ambient_2017}, is still the most probable structure for bulk high pressure phase to \SI{145}{GPa} as the peak positions best line up with the experimental XRD patterns.
The extra peaks  in Fig.~\ref{WB2_HPXRD}e, which are marked by the red and black asterisk symbols, come from the Ne pressure medium and Re gasket, respectively. To account for the effects of stress and strain in \ch{WB2} under nonhydrostatic pressure condition similar to the electrical resistivity measurements, we have performed XRD measurements up to \SI{98}{GPa} without any pressure medium filling the \ch{Re} gasket hole with only the \ch{WB2} sample (See Supplementary Figure 9). The results indicate mostly hP12 phase except for one or two peaks appearing above \SI{50}{GPa} that may be due to hP3 or hR6 phase as the possible local defects. Therefore, it is concluded that there is no bulk structural transition to hP3 or hR6 structure and \ch{WB2} remains predominantly in the hP12 phase even in nonhydrostatic pressure condition in agreement with the quasi-hydrostatic XRD measurements in Fig.~\ref{WB2_HPXRD}.
Figure~\ref{WB2_HPXRD}f show the crystal structures of five competing phases of \ch{WB2} from Zhang {\it et al.}~\cite{Zhang2017}. Out of these five competing phases, the tI12 phase forms a 3-d network structure. The remaining four competing phases form a layered structure, where the boron layers are either buckled or unbuckled depending on the position of the tungsten atoms in between the layers.

\subsection{Theory of bulk phases}
\begin{figure}[ht]
  \includegraphics[width=0.8\columnwidth]{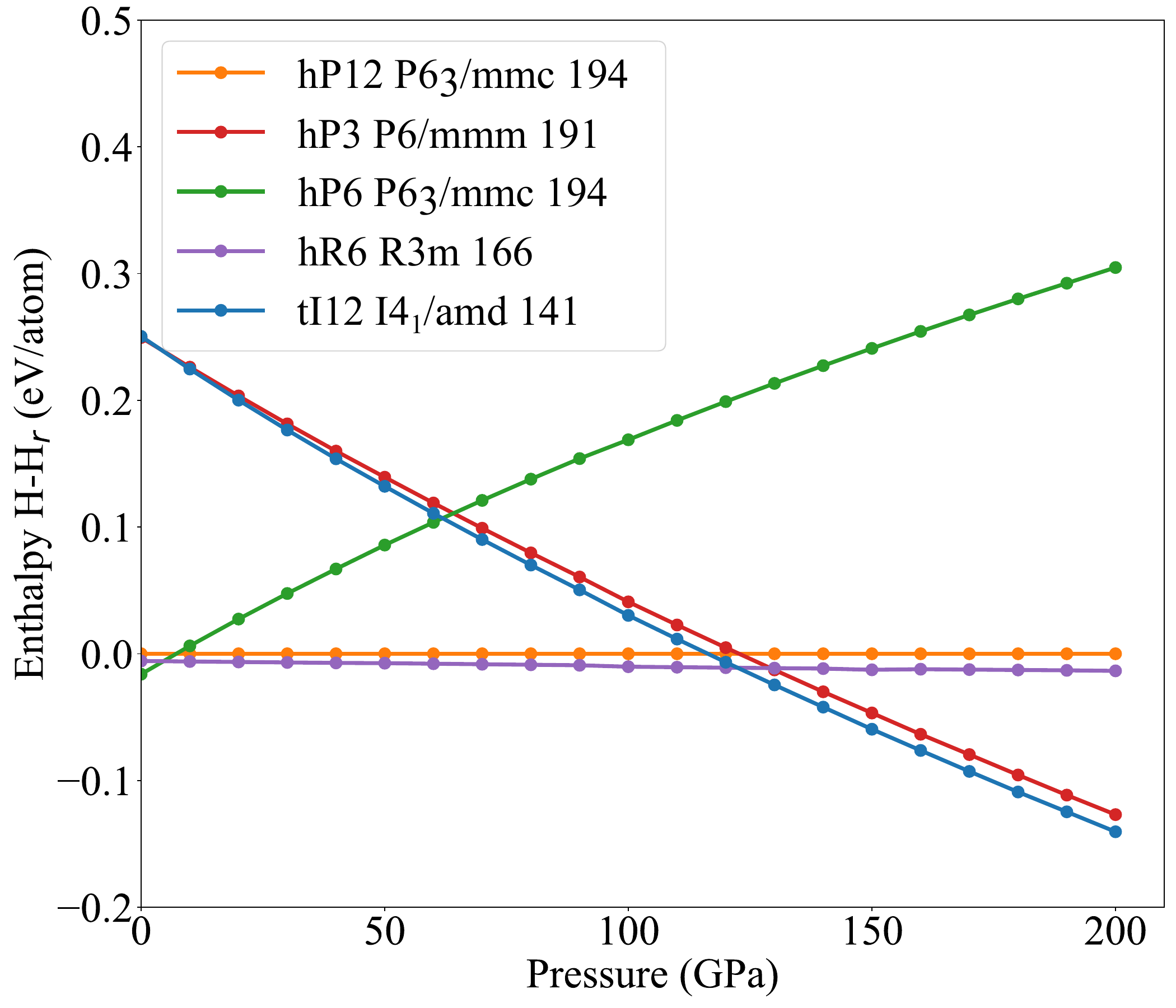}
  \caption{Enthalpy of various competing phases of \ch{WB2} as a function of pressure. The hP6 phase of \ch{WB2} has the lowest enthalpy at ambient pressure, and the enthalpies of hP12 and hR6 structures are higher only 16 and 7 meV/atom, respectively.}
  \label{WB2_enthalpy}
\end{figure}
Figure \ref{WB2_enthalpy} shows the calculated enthalpy as a function of pressure for various competing phases of \ch{WB2}.
According to our calculations, the \ch{WB2} hP6 structure has (by a small margin) the lowest enthalpy at ambient pressure conditions but is not observed in the experimental sample.
The theoretical enthalpy of the experimentally observed hP12 structure is about $\sim\SI{16}{meV/atom}$ higher than the hP6 structure. We used arc melting to synthesize the samples, which were then quickly cooled to room temperatures on a water cooled Cu hearth. At the arc melting temperatures ($\sim$2370 K), entropic contributions to the Gibbs free energy can easily overcome the energy difference of 16 meV/atom between the hP12 and hP6 phases. And because of the quick cooling of the samples, the high-temperature metastable hP12 phase, as is evident from the XRD, is retained at room temperature. At ambient pressure and low temperature, one can expect a phase transformation from the metastable hP12 phase to the DFT ground state hP6 phase, likely via nucleation and growth. However, due to the vanishingly small diffusivity at low temperatures, such a phase transformation is unlikely. Moreover, with the accuracy of the existing exchange-correlation functional, the difference in energy is too small to resolve the question of the 'true' low-temperature ground state \cite{Zhang2018,Yang2019,Hinuma2017,Bartel2019}. Even if the DFT prediction of metastability of 16 meV/atom were correct, we expect the hP12 phase to remain sufficiently metastable.

The hP12 and hR6 structures both have equal number of planar and buckled B layers in their respective unit cells and are related by changes in stacking sequences of W planes along the $c$-axis.
If the stacking sequence of W atoms in hP12 is labeled as `AA-BB-AA-BB' then the the stacking sequence in hR6 is `AA-BB-CC-AA-BB-CC'.
This is also evidenced by the nature of the enthalpy vs pressure curves of the two phases. With increasing pressure, the stability of the hP12 structure increases.
From Fig.~\ref{WB2_HPXRD}f, the hP6 phase has no planar boron layers. Empirically, the lack of planar B layers might be responsible for the increasing enthalpy of the hP6 phase with pressure. The tI12 phase consists of B layers with a 90-degree twist at every c/4 increment  along the c-axis.
In contrast to the hP6 phase, the hP3 phase has only planar B layers. Above $\sim \SI{130}{GPa}$ one can expect the phase transition hP12 $\rightarrow$ hP3, but kinetic barriers may prevent this phase transition.

To investigate the origin of pressure-induced superconductivity in this material, we performed electron-phonon calculations under pressure to determine the theoretical electron-phonon superconducting critical temperature for the hP3,  hP12, and hR6 phases.
Table~\ref{tab:a2F} summarizes the electron-phonon coupling strength $\lambda$ and the frequency moments $\bar\omega_2$ and $\omega_{\mathrm log}$ obtained from the Quantum Espresso code.
Using the Allen-Dynes equation with $\mu^\ast = 0.13$, or the formula by Xie {\it et al}~\cite{Xie2021},  we  find that while both the hR6 and the hP12 structure have consistently subkelvin $T_c$'s up to 140~GPa, the hP3 structure has a critical temperature of 30 K, relatively insensitive to pressure up to 140 GPa.
If experimental lattice constants are used (same $a$ value but four times smaller $c$ in hP12 at 145 GPa), this critical temperature is found to be lower, about 20~K.
Details are presented in the Methods section. 

It is difficult at first sight to reconcile these results with our data.
A structural transition from hP12 to hP3 around 50 GPa would be qualitatively consistent with the $T_c$ data, but there are no clear signatures of hP3 lines in the XRD analysis at any pressure.
Furthermore, the theoretical enthalpy difference between hP12
and hP3 phases around 50 GPa is too large to allow the hP12 $\rightarrow$ hP3 bulk phase transformation.
Calculations of the density of states near the Fermi surface and the electron-phonon coupling in the hP12 state, on the other hand, show no dramatic changes with pressure, and cannot explain the jump in $T_c$ at 70 GPa with this structure alone.
\begin{table}[t]
    \centering
    \caption{Pressure (P in GPa), the $ \omega_{\log} $ and $ \bar\omega_2$ moments of the spectral function $\alpha^2F(\omega)$ in units of meV and the predicted superconducting transition temperature $T_c$ using the Allen-Dynes equation~\cite{Allen-Dynes1975}, isotropic Eliashberg, and the machine learning equation by Xie {\it et al.} ~\cite{Xie2021} with $\mu^\ast = 0.13$.}
    \label{tab:a2F}
    \begin{ruledtabular}
    \begin{tabular}{c c c c c c c c}
        Phase & $P$ & $\lambda$ & $ \omega_{\log}$ (meV) & $ \bar\omega_2$ (meV) & $T_c^\mathrm{AD}$ & $T_\mathrm{c}^\mathrm{El}$ & $T_\mathrm{c}^\mathrm{Xie}$ (K) \\
        \colrule
         hP12  & 0 & 0.37 & 30.3 & 47.2 & 0.4 & - & 0.4  \\
        hP12  & 100 & 0.29 & 41.5& 63.6& 0 & -& 0 \\
        hP3   & 100 & 1.72 & 17.3& 35.5& 27.8 &  27.3 & 34.5\\
        hR6   & 0   & 0.53  & 27.2 & 43.2 & 3.1 &-  & 2.8 \\
        hR6 & 100 & 0.38 & 40.4 & 61.8 & 0.66 & - & 0.6\\
    \end{tabular}
    \end{ruledtabular}
\end{table}

\section{Discussion}
In their discovery of superconductivity in \ch{MoB2}, Pei et al.~\cite{MoB2_superconductivity} stress that their theoretical calculations indicate important roles for both the Mo d-electrons and for the phonon modes of the Mo.  By contrast, in MgB$_2$ B phonons dominate the electron-phonon coupling strength~\cite{Mazin2003}.
Under pressure, Pei et al.~\cite{MoB2_superconductivity} find that the low pressure $\beta$ structure of \ch{MoB2} (hR6, space group 166,  R$\bar{3}$m, structure prototype \ch{CaSi2}) transforms to the high pressure $\alpha$ \ch{MoB2} (hP3, space group 191, P6/mmm, structure prototype AlB$_2$) structure, i.e.,\ the same structure as \ch{MgB2}.

For \ch{WB2}, Fig.~\ref{WB2_HPXRD}d shows the experimental XRD pattern in compared with those calculated theoretically for structures relaxed using DFT with PBEsol functional.
At high pressures, the peaks from the competing phases line up at certain angles, making it difficult to determine the correct phase from the XRD peaks alone.
The most likely crystal structure at high pressures can be inferred by combining information from Fig.~\ref{WB2_enthalpy}, Fig.~\ref{WB2_HPXRD}, and the theoretically calculated superconducting critical temperatures of various competing phases at high pressure.
The tI12 and hP6 structures of \ch{WB2} can be eliminated as the most likely structures at high pressure, since both tI12 and hP6 structures have high theoretical enthalpy, and the XRD peaks of both tI12 and hP6 do not match the experimentally observed peaks.

As further evidence of a lack of bulk structural phase transition in our samples,
the electrical resistivity curves of our sample shown in Fig.~\ref{fig:WB2_rhoPT}c initially decrease  monotonically with pressure and show no clear signature of structural transformation.
Because the DOS and electron-phonon coupling show no specific features occurring at or near 50 GPa, we believe that the initial decrease can be attributed to a hardening of the phonon spectrum which  reduces the scattering phase space, together with a weak reduction of the electronic DOS with pressure.  Both of these effects are indeed seen in our calculations. The minimum must therefore occur because of a relatively rapid increase in scattering in the sample around 50 GPa, of unknown origin. 

The same resistivity argument can also help rule out the bulk structural transition to the hR6 phase.
In the high-pressure XRD, some of the peaks corresponding to the hR6 phase are missing.
For example, at 145 GPa the theoretical peaks of the hR6 phase at  $\approx 8.8^{\circ}$  (i.e., peak by ($10\bar{1}5$) plane) and $\approx 9.9^{\circ}$ (i.e peak by (10$\bar{1}$7) plane) are missing from the experimentally measured peaks. Can a strained hR6  lattice produce the experimentally measured peaks? To answer this question we artificially strain the 'a' lattice parameter of our DFT relaxed structure and calculate the XRD pattern. The 'a' lattice parameter was strained by $\pm 7\%$ in steps of $2\%$ as compared to the relaxed structure. The 'c' lattice parameter of our structure was kept unchanged as the XRD peaks that have contributions only by the planes along the 'c' direction of the hR6 phase line-up almost perfectly with the experimentally measured peaks. Supplementary Figures 1 and 2 shows the theoretical and experimental XRD peaks of the strained hR6 phase at 85 and 145~GPa.
The peaks of the strained hR6 structures fail to produce a diffraction pattern that is in agreement with the one measured experimentally.
Even if the hR6 phase is present in our sample at high pressures it could be well below the detection limit of the XRD apparatus.
Can the presence of a very small amount of the hR6 phase, undetectable by XRD, account for the superconductivity seen in our samples at high pressure?
From the calculated critical temperature in Table~\ref{tab:a2F} high pressure superconductivity deriving from the hR6 phase is not a credible explanation.

We therefore propose that the superconductivity onset at \SI{55}{GPa} in our samples is due to a filamentary phase formed from stacking faults known to occur in this system~\cite{Kahyan2012}.
Since our calculations show that the MgB$_2$-like hP3 phase has a high critical temperature, it 
appears likely that as the material is plastically deformed with increasing pressure, stacking faults and twin boundaries form.
This scenario is further supported by the metastable superconducting behavior of \ch{WB2} during decompression (See Fig.~\ref{fig:WB2_rhoPT}d).
As discussed below, the structure of these defects can resemble either the structure of the hR6 or hP3 phase locally.
A scenario that is consistent with all of the computational and experimental evidence is that the concentration of hP3 defects increases sharply at around \SI{50}{GPa}, leading to a rise in the resistivity, and, eventually, a percolating path for superconductivity formed around these defects.

Defects of the hR6-type structure in the ambient pressure polycrystalline sample of Ref.~\cite{Kahyan2012} may also explain the observation of a $T_c$ of a few K.
It is interesting to note that when the grains of the samples fabricated in that work decreased in size, superconductivity disappeared.
This is consistent with the ease with which defects can migrate to a grain boundary in smaller crystallites.
It is also suggests that our single crystal sample contains many fewer stacking faults at ambient pressure, such that defect-induced superconductivity does not occur.

\begin{figure*}
    \centering
    \includegraphics[width=0.8\linewidth]{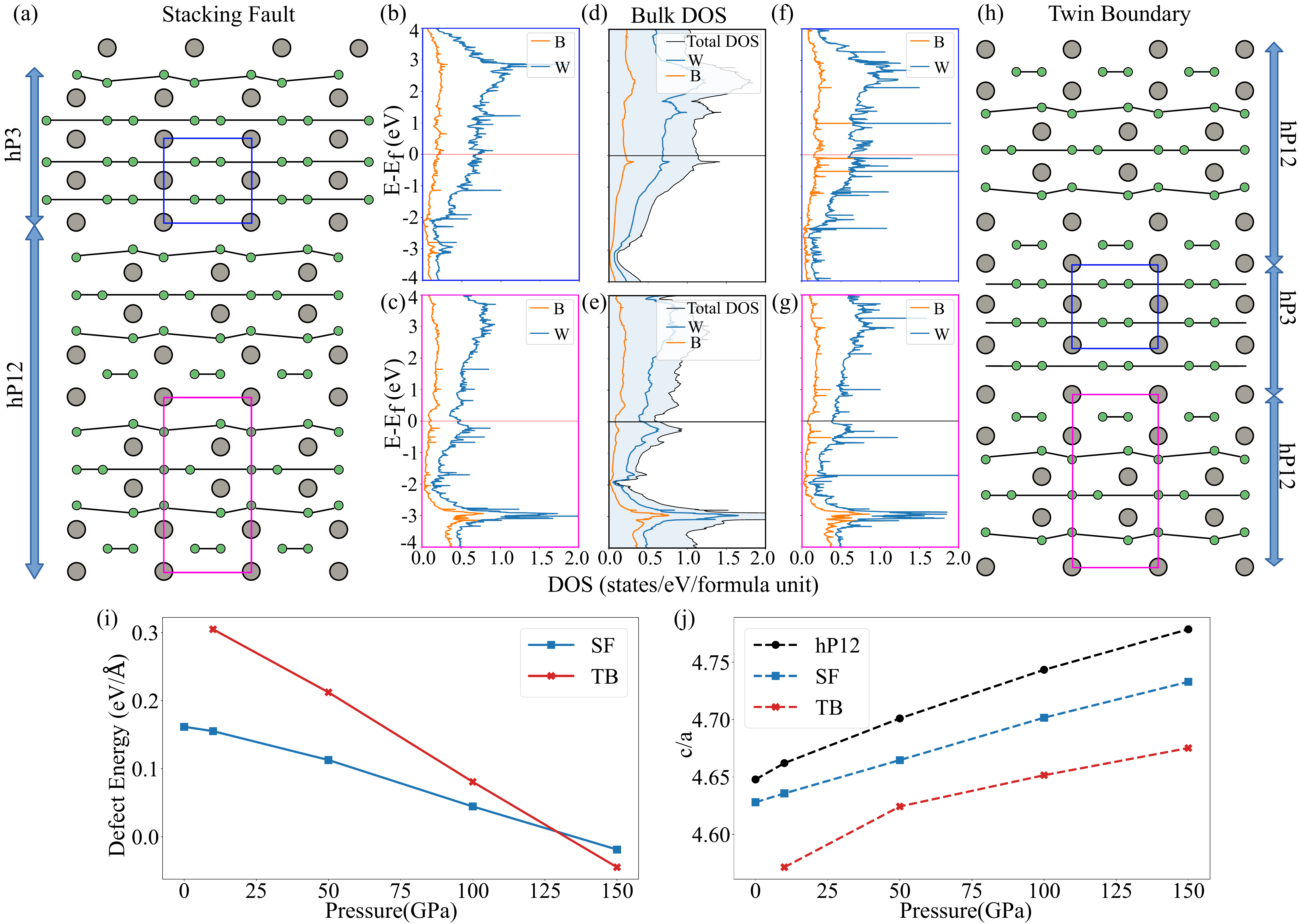}
    \caption{(a) Stacking faults and (e) Twin boundaries in the hP12 phase; the stacking of atomic planes along the c-axis at the defect mirrors the stacking in the hP3 phase. Green and gray circles represent boron and tungsten respectively. (b)-(c), and (f)-(g) projected DOS for the stacking faults and twin boundary. The DOS was spatially projected on the hP3 and hP12 parts of the defect structures. (d)-(e) DOS of the bulk hP3 and hP12 phases. All the DOS were calculated at 50 GPa in VASP with PBEsol for exchange-correlation energy. (i) Stacking fault, and twin boundary energy for the hP12 phase as a function of pressure. (j) Theoretical $c/a$ ratio for the hP12 phase with planar defects as a function of pressure. The red curve (crosses) and the blue curve (squares) represent the twin boundary (TB) and the stacking faults (SF) respectively.}
    \label{WB2_SF_TB}
\end{figure*}
{Supplementary Figures 3 and 4 show how stacking fault and twin boundary defects can be formed in the hP12 phase by sliding appropriate planes. With the introduction of stacking faults and twin boundaries in the hP12 phase, the local environment at the defects can become similar to the hP3 phase. 

Figures \ref{WB2_SF_TB}(a) and \ref{WB2_SF_TB}(h) shows the hP3 region formed because of stacking faults and twin boundaries, respectively. Figure \ref{WB2_SF_TB}(i) shows the calculated stacking fault and twin boundary energy, given by 
\begin{equation}
    FE_{D} = \frac{H_D - (N_D*H_{\text{hP12}})}{A},
    \label{sf_tb}
\end{equation}
where $FE_{D}$ is the formation energy of the defect, $H_D$ is the enthalpy of the defect structure, $N_D$ are the number of atoms in the defect structure simulation cell, $H_{\text{hP12}}$ is the enthalpy of the hP12 phase and $A$ is the area. The VASP DFT code with PBEsol functional for exchange-correlation was used to calculate the enthalpy of the defect structures. A k-point density of $\SI{60}{/\angstrom}$ and a cut-off energy of $\SI{520}{\electronvolt}$ for the plane wave basis set were used in the calculation. The simulation cell used for both the defects were made out of 4 hP12 unit cells stacked on top of each other.
Both the stacking fault and twin boundary energy decrease as pressure increases, enhancing the probability of formation of these defects.
Figure \ref{WB2_SF_TB}(b) and \ref{WB2_SF_TB}(c) show the density of states (DOS) projected onto the atoms in the hP12 and the hP3 regions of the stacking fault defect structure calculated at $\SI{50}{GPa}$. These spatially projected DOS closely match the DOS of the pure bulk hP12 and hP3 phases, respectively. From these figures \ref{WB2_SF_TB}(a)-(h), it is clear that both the atomic and electronic structures of the stacking fault match those of the pure high $T_c$ hP3 phase.
Similar correlations can also be seen between the projected DOS of the twin boundary (Fig.~\ref{WB2_SF_TB}(f) and \ref{WB2_SF_TB}(g)) and that of the pure phases. Both the structural and electronic similarity between the atoms at the planar defect structures and the pure hP3 phase strongly point towards the possibility of superconductivity because of these planar defects.}
Figure \ref{WB2_SF_TB}(j) shows the theoretical c/a ratio of the stacking fault and twin boundary defect structures compared to that of a perfect hP12 structure. Thus, one might expect that at a given pressure, the introduction of planar defects in the hP12 phase will cause the c/a ratio to decrease. Unfortunately, such a reduction might be below the resolution of XRD apparatus to be observed experimentally and as shown in Fig.~\ref{WB2_HPXRD} the theoretical c/a ratio of the hP12 phase and the experimental c/a ratio match quite well.

Figure \ref{WB2_SF_TB}(j) shows the theoretical $c/a$ ratio of the hP12 phase with stacking fault and twin boundary defects compared to that of a perfect hP12 structure. Thus, at a given pressure the introduction of planar defects in the hP12 phase should cause the $c/a$ ratio to decrease.
Just such a reduction in the $c/a$ ratio at $\sim$\SI{60}{GPa} from its low pressure extrapolated values is also observed experimentally in the inset of Fig.~\ref{WB2_HPXRD}(c), providing further evidence for formation of planar defects with increasing pressure.

In summary, we measured the resistive transition of \ch{WB2} crystals under pressure up to 187 GPa, and shown that superconductivity around 17 K begins near 80 GPa, and that this evolution takes place without a bulk structural transition in the sample.
According to x-ray analysis, the system  remains nearly entirely in the same bulk crystal structure as at ambient pressure through the onset of $T_c$.
None of the other competing bulk crystal structures are close enough in enthalpy to form, nor do they appear in the XRD patterns.
The results lead to the fascinating and plausible scenario in which defects that resemble the hP3 MgB$_2$ structure locally carry the superconductivity, but are present in filamentary quantities only.
This appears to be a novel way to create superconductivity under pressure, and may point to a path to lower the critical pressure of high-$T_c$ superconductors like the hydrides currently under intense investigation.

\section{Methods}
Boron pieces (99.98\% pure) were wrapped in 99.9\% pure tungsten sheet in stoichiometric amounts and arc-melted together.
Despite the high melting point (\SI{2400}{\degree C}) of \ch{WB2}, the low vapor pressures of both B and W at this temperature led to negligible mass loss upon melting the constituents together and remelting twice.
Upon cooling, the arc-melted bead showed hexagonal crystal facets on the surface (see Fig. 2d), pieces of which were harvested for the high pressure  measurements.
For an example of this method of single crystal production, see Ref.~\cite{Heuser2000}.

\subsection{High pressure methods}
High pressure x-ray diffraction measurements were performed on a powdered piece of a single crystal facet from the arc melted button of \ch{WB2} at Argonne National Laboratory's Advanced Photon Source, beamline 16-BM-D.
The x-ray beam had a wavelength of $\SI{0.41}{\angstrom}$ ($\SI{30}{\kilo\electronvolt}$) in run 1 and $\SI{0.31}{\angstrom}$ ($\SI{40}{\kilo\electronvolt}$) in run 2, respectively. The x-ray beam was focused to a $\sim$ $\SI{5}{\micro\meter}$ by $\SI{5}{\micro\meter}$ (FWHM) spot at the sample.
An MAR345 image plate detector was used to record the diffracted intensity.
Exposure times were typically $\sim$ 60 to 120 seconds per image.
A \ch{CeO2} standard was used to calibrate the sample to detector distance.
Neon was used as the pressure medium.
Pressure was determined both using an online ruby fluorescence measurement as well as the equation of state of Au grains loaded into the sample chamber.
DIOPTAS~\cite{Dioptas2015} software was used to convert the 2D diffraction images to 1D diffraction patterns.
The resulting XRD patterns were then further analyzed by Rietveld~\cite{Rietveld_1969} or Le Bail~\cite{LEBAIL1988} methods using GSAS-II software~\cite{Toby_GSASII_2013}.
The visualization of the crystal structure was depicted using VESTA software~\cite{Momma_VESTA_2008}.

For the high-pressure resistivity measurements, a micron-sized \ch{WB2} single crystal sample ($\sim$ 40 $\times$ 40 $\times$ $\SI{10}{\micro\m^3}$) was placed in a gas-membrane-driven diamond anvil cell (OmniDAC from Almax-EasyLab) along with a ruby ($\sim \SI{20}{\micro\m}$ in diameter) for pressure calibration~\cite{chijioke_ruby_2005} below \SI{100}{GPa}, above which diamond anvil Raman was used~\cite{Akahama2006} at $\sim$\SI{10}{K}.
Pressure was determined via automatic real-time fitting of the ruby spectrum, allowing dense sampling of resistance versus pressure, as the load was smoothly adjusted using a Pace 5000 computer-controlled pressure regulator.
Two opposing diamond anvils (type Ia, 1/6-carat, \SI{0.15}{\milli\m} central flats) were used. A Re metal gasket was pre-indented from $\sim$ 250 to $\SI{26}{\micro\m}$ in thickness with a hole ($\sim \SI{140}{\micro\m}$ in diameter), which was filled with a 4:1 cBN-epoxy mixture and soapstone (relatively soft) for outer and inner areas, respectively, to electrically insulate the sample from the metal gasket and also serving as the pressure-transmitting medium (see insets in Fig.~\ref{fig:WB2_rhoPT}c). The thin \ch{WB2} sample was then placed on top of four thin and pointy Pt leads ($\sim$ \SI{4}{\micro\meter} thick), which were extended by other four longer Pt leads, for a four-point dc electrical resistivity measurement. Further details of the nonhydrostatic high pressure resistivity technique are given in a paper by Matsuoka $et$ $al.$~\cite{Matsuoka2009}.

The diamond cell was placed inside a customized continuous-flow cryostat (Oxford Instruments). A home-built optical system attached to the bottom of the cryostat was used for the visual observation of the sample and for the measurement of the ruby manometer. Pressure was applied at room temperature to the desired pressure, and then the sample was cooled down to $\sim \SI{1.8}{K}$ and warmed up to room temperature at a rate of $\sim \SI{0.25}{K/min}$ at each pressure for the temperature-dependent resistivity measurement. To estimate the electrical resistivity from the resistance, we used the van der Pauw method, (assuming an isotropic sample in the measurement plane), ${\rho} = {\pi}tR/\ln{2}$, where $t$ is the sample thickness ($\sim\SI{10}{\micro\meter}$) with currents of $0.1-\SI{1}{\milli\ampere}$. The accuracy of the estimated resistivity is roughly a factor of two or three considering uncertainties in the initial thickness of the sample. No attempt was made to take into account the changes in the sample thickness under high pressures.
For the upper critical field measurements, we used a Quantum Design physical property measurement system (PPMS) and an Almax-EasyLab Chicago Diamond Anvil Cell (Chicago-DAC) with two opposing diamond anvils (0.15 and 0.5 mm central flats), whose ruby pressure was measured at room temperature and estimated for the small change in pressure at $\sim$\SI{10}{K}. One of the diamonds was a designer-diamond anvil (0.15 mm central flat) with six symmetrically deposited tungsten microprobes in the encapsulated high-quality-homoepitaxial diamond~\cite{weir_epitaxial_2000}.

\subsection{Computational methods}
To investigate possible phase transitions, at high pressure we calculated the enthalpy as a function of pressure for the various competing phases of \ch{WB2}. The structures of competing phases were obtained from Zhang {\it et al.}~\cite{Zhang2017}, who investigated the crystal structures of \ch{WB2} up to 200 GPa using the particle swarm optimization algorithm. We used VASP~\cite{Kresse1996,Kresse1996_2} with the PBEsol functional~\cite{Csonka2009} for the exchange-correlation energy~\cite{Perdew2008} along with the projector augmented wave (PAW) pseudopotentials\cite{Blchl1994} for structural relaxation. A plane wave cut-off of \SI{520}{\electronvolt} and a k-point density of $\SI{60}{/\angstrom}$ are used in the calculation. The Methfessel-Paxton method was used for smearing the electrons near the Fermi level with a smearing value of \SI{0.1}{\electronvolt}~\cite{Methfessel1989}. For the DOS calculations we used the tetrahedron method with Bl\"ochl correction~\cite{Blchl1994_paw}.

Electron-phonon coupling calculations for several phases of WB$_2$ are carried out using
the linear response method as implemented in the Quantum Espresso code~\cite{qe1, qe2, qe3}.
The exchange correlation potential is chosen to be PBE \cite{PBE} and we have used
the optimized norm-conserving pseudopotential\cite{Hamann2013,Schlipf2015}. 
The wave function cutoff is set to 60 Ry and the 
charge density cutoff is fixed at 240 Ry. For the hP12 and hR6 phases, the k-mesh consists of 16$\times$16$\times$16 points in the
whole Brillouin zone to preserve crystal symmetry, and the q-mesh is 4$\times$4$\times$4. Brillouin zone integration was carried out using the optimized tetrahedron method~\cite{PhysRevB.89.094515}.
For the hP3 (AlB$_2$) phase, we first calculate the phonon dispersion on  coarse k and q-meshes with $18\times18\times18$ and $6\times6\times6$ points, respecitvely, which are later interpolated onto fine k and q-meshes with $60\times60\times60$ and $30\times30\times30$ points respectively.
Isotropic Eliashberg equations are solved to obtain the transition temperatures of the hP3 (AlB$_2$) phase under pressure.

\vspace{1em}
\noindent
\textbf{Data availability statement:}
The datasets generated and/or analysed during the current study are available from the corresponding author on reasonable request.

\bibliography{WB2}

\begin{thebibliography}{47}%
\makeatletter
\providecommand \@ifxundefined [1]{%
 \@ifx{#1\undefined}
}%
\providecommand \@ifnum [1]{%
 \ifnum #1\expandafter \@firstoftwo
 \else \expandafter \@secondoftwo
 \fi
}%
\providecommand \@ifx [1]{%
 \ifx #1\expandafter \@firstoftwo
 \else \expandafter \@secondoftwo
 \fi
}%
\providecommand \natexlab [1]{#1}%
\providecommand \enquote  [1]{``#1''}%
\providecommand \bibnamefont  [1]{#1}%
\providecommand \bibfnamefont [1]{#1}%
\providecommand \citenamefont [1]{#1}%
\providecommand \href@noop [0]{\@secondoftwo}%
\providecommand \href [0]{\begingroup \@sanitize@url \@href}%
\providecommand \@href[1]{\@@startlink{#1}\@@href}%
\providecommand \@@href[1]{\endgroup#1\@@endlink}%
\providecommand \@sanitize@url [0]{\catcode `\\12\catcode `\$12\catcode
  `\&12\catcode `\#12\catcode `\^12\catcode `\_12\catcode `\%12\relax}%
\providecommand \@@startlink[1]{}%
\providecommand \@@endlink[0]{}%
\providecommand \url  [0]{\begingroup\@sanitize@url \@url }%
\providecommand \@url [1]{\endgroup\@href {#1}{\urlprefix }}%
\providecommand \urlprefix  [0]{URL }%
\providecommand \Eprint [0]{\href }%
\providecommand \doibase [0]{http://dx.doi.org/}%
\providecommand \selectlanguage [0]{\@gobble}%
\providecommand \bibinfo  [0]{\@secondoftwo}%
\providecommand \bibfield  [0]{\@secondoftwo}%
\providecommand \translation [1]{[#1]}%
\providecommand \BibitemOpen [0]{}%
\providecommand \bibitemStop [0]{}%
\providecommand \bibitemNoStop [0]{.\EOS\space}%
\providecommand \EOS [0]{\spacefactor3000\relax}%
\providecommand \BibitemShut  [1]{\csname bibitem#1\endcsname}%
\let\auto@bib@innerbib\@empty
\bibitem [{\citenamefont {Tomita}\ \emph {et~al.}(2001)\citenamefont {Tomita},
  \citenamefont {Hamlin}, \citenamefont {Schilling}, \citenamefont {Hinks},\
  and\ \citenamefont {Jorgensen}}]{tomita_dependence_2001}%
  \BibitemOpen
  \bibfield  {author} {\bibinfo {author} {\bibfnamefont {T.}~\bibnamefont
  {Tomita}}, \bibinfo {author} {\bibfnamefont {J.~J.}\ \bibnamefont {Hamlin}},
  \bibinfo {author} {\bibfnamefont {J.~S.}\ \bibnamefont {Schilling}}, \bibinfo
  {author} {\bibfnamefont {D.~G.}\ \bibnamefont {Hinks}}, \ and\ \bibinfo
  {author} {\bibfnamefont {J.~D.}\ \bibnamefont {Jorgensen}},\ }\href {\doibase
  10.1103/PhysRevB.64.092505} {\bibfield  {journal} {\bibinfo  {journal}
  {Physical Review B}\ }\textbf {\bibinfo {volume} {64}},\ \bibinfo {pages}
  {092505} (\bibinfo {year} {2001})},\ \bibinfo {note} {publisher: American
  Physical Society}\BibitemShut {NoStop}%
\bibitem [{\citenamefont {Deemyad}\ \emph {et~al.}(2003)\citenamefont
  {Deemyad}, \citenamefont {Tomita}, \citenamefont {Hamlin}, \citenamefont
  {Beckett}, \citenamefont {Schilling}, \citenamefont {Hinks}, \citenamefont
  {Jorgensen}, \citenamefont {Lee},\ and\ \citenamefont
  {Tajima}}]{deemyad_dependence_2003}%
  \BibitemOpen
  \bibfield  {author} {\bibinfo {author} {\bibfnamefont {S.}~\bibnamefont
  {Deemyad}}, \bibinfo {author} {\bibfnamefont {T.}~\bibnamefont {Tomita}},
  \bibinfo {author} {\bibfnamefont {J.~J.}\ \bibnamefont {Hamlin}}, \bibinfo
  {author} {\bibfnamefont {B.~R.}\ \bibnamefont {Beckett}}, \bibinfo {author}
  {\bibfnamefont {J.~S.}\ \bibnamefont {Schilling}}, \bibinfo {author}
  {\bibfnamefont {D.~G.}\ \bibnamefont {Hinks}}, \bibinfo {author}
  {\bibfnamefont {J.~D.}\ \bibnamefont {Jorgensen}}, \bibinfo {author}
  {\bibfnamefont {S.}~\bibnamefont {Lee}}, \ and\ \bibinfo {author}
  {\bibfnamefont {S.}~\bibnamefont {Tajima}},\ }\href {\doibase
  10.1016/S0921-4534(02)02300-6} {\bibfield  {journal} {\bibinfo  {journal}
  {Physica C: Superconductivity}\ }\textbf {\bibinfo {volume} {385}},\ \bibinfo
  {pages} {105} (\bibinfo {year} {2003})}\BibitemShut {NoStop}%
\bibitem [{\citenamefont {Buzea}\ and\ \citenamefont
  {Yamashita}(2001)}]{buzea_review_2001}%
  \BibitemOpen
  \bibfield  {author} {\bibinfo {author} {\bibfnamefont {C.}~\bibnamefont
  {Buzea}}\ and\ \bibinfo {author} {\bibfnamefont {T.}~\bibnamefont
  {Yamashita}},\ }\href {\doibase 10.1088/0953-2048/14/11/201} {\bibfield
  {journal} {\bibinfo  {journal} {Superconductor Science and Technology}\
  }\textbf {\bibinfo {volume} {14}},\ \bibinfo {pages} {R115} (\bibinfo {year}
  {2001})}\BibitemShut {NoStop}%
\bibitem [{\citenamefont {Bud’ko}\ and\ \citenamefont
  {Canfield}(2015)}]{budko_superconductivity_2015}%
  \BibitemOpen
  \bibfield  {author} {\bibinfo {author} {\bibfnamefont {S.~L.}\ \bibnamefont
  {Bud’ko}}\ and\ \bibinfo {author} {\bibfnamefont {P.~C.}\ \bibnamefont
  {Canfield}},\ }\href {\doibase 10.1016/j.physc.2015.02.024} {\bibfield
  {journal} {\bibinfo  {journal} {Physica C: Superconductivity and its
  Applications}\ }\bibinfo {series} {Superconducting {Materials}:
  {Conventional}, {Unconventional} and {Undetermined}},\ \textbf {\bibinfo
  {volume} {514}},\ \bibinfo {pages} {142} (\bibinfo {year}
  {2015})}\BibitemShut {NoStop}%
\bibitem [{\citenamefont {Pei}\ \emph {et~al.}(2021)\citenamefont {Pei},
  \citenamefont {Zhang}, \citenamefont {Wang}, \citenamefont {Zhao},
  \citenamefont {Gao}, \citenamefont {Gong}, \citenamefont {Tian},
  \citenamefont {Luo}, \citenamefont {Lu}, \citenamefont {Lei}, \citenamefont
  {Liu},\ and\ \citenamefont {Qi}}]{MoB2_superconductivity}%
  \BibitemOpen
  \bibfield  {author} {\bibinfo {author} {\bibfnamefont {C.}~\bibnamefont
  {Pei}}, \bibinfo {author} {\bibfnamefont {J.}~\bibnamefont {Zhang}}, \bibinfo
  {author} {\bibfnamefont {Q.}~\bibnamefont {Wang}}, \bibinfo {author}
  {\bibfnamefont {Y.}~\bibnamefont {Zhao}}, \bibinfo {author} {\bibfnamefont
  {L.}~\bibnamefont {Gao}}, \bibinfo {author} {\bibfnamefont {C.}~\bibnamefont
  {Gong}}, \bibinfo {author} {\bibfnamefont {S.}~\bibnamefont {Tian}}, \bibinfo
  {author} {\bibfnamefont {R.}~\bibnamefont {Luo}}, \bibinfo {author}
  {\bibfnamefont {Z.-Y.}\ \bibnamefont {Lu}}, \bibinfo {author} {\bibfnamefont
  {H.}~\bibnamefont {Lei}}, \bibinfo {author} {\bibfnamefont {K.}~\bibnamefont
  {Liu}}, \ and\ \bibinfo {author} {\bibfnamefont {Y.}~\bibnamefont {Qi}},\
  }\href@noop {} {\enquote {\bibinfo {title} {Pressure-induced
  superconductivity at 32 k in mob2},}\ } (\bibinfo {year} {2021}),\ \Eprint
  {http://arxiv.org/abs/arXiv:2105.13250} {arXiv:2105.13250} \BibitemShut
  {NoStop}%
\bibitem [{\citenamefont {Kayhan}\ \emph {et~al.}(2012)\citenamefont {Kayhan},
  \citenamefont {Hildebrandt}, \citenamefont {Frotscher}, \citenamefont
  {Senyshyn}, \citenamefont {Hofmann}, \citenamefont {Alff},\ and\
  \citenamefont {Albert}}]{Kahyan2012}%
  \BibitemOpen
  \bibfield  {author} {\bibinfo {author} {\bibfnamefont {M.}~\bibnamefont
  {Kayhan}}, \bibinfo {author} {\bibfnamefont {E.}~\bibnamefont {Hildebrandt}},
  \bibinfo {author} {\bibfnamefont {M.}~\bibnamefont {Frotscher}}, \bibinfo
  {author} {\bibfnamefont {A.}~\bibnamefont {Senyshyn}}, \bibinfo {author}
  {\bibfnamefont {K.}~\bibnamefont {Hofmann}}, \bibinfo {author} {\bibfnamefont
  {L.}~\bibnamefont {Alff}}, \ and\ \bibinfo {author} {\bibfnamefont
  {B.}~\bibnamefont {Albert}},\ }\href {\doibase
  https://doi.org/10.1016/j.solidstatesciences.2012.05.036} {\bibfield
  {journal} {\bibinfo  {journal} {Solid State Sciences}\ }\textbf {\bibinfo
  {volume} {14}},\ \bibinfo {pages} {1656} (\bibinfo {year} {2012})},\ \bibinfo
  {note} {the 17th International Symposium on Boron, Borides and Related
  Materials}\BibitemShut {NoStop}%
\bibitem [{\citenamefont {Frotscher}\ \emph {et~al.}(2007)\citenamefont
  {Frotscher}, \citenamefont {Klein}, \citenamefont {Bauer}, \citenamefont
  {Fang}, \citenamefont {Halet}, \citenamefont {Senyshyn}, \citenamefont
  {Baehtz},\ and\ \citenamefont {Albert}}]{Frotscher2007}%
  \BibitemOpen
  \bibfield  {author} {\bibinfo {author} {\bibfnamefont {M.}~\bibnamefont
  {Frotscher}}, \bibinfo {author} {\bibfnamefont {W.}~\bibnamefont {Klein}},
  \bibinfo {author} {\bibfnamefont {J.}~\bibnamefont {Bauer}}, \bibinfo
  {author} {\bibfnamefont {C.-M.}\ \bibnamefont {Fang}}, \bibinfo {author}
  {\bibfnamefont {J.-F.}\ \bibnamefont {Halet}}, \bibinfo {author}
  {\bibfnamefont {A.}~\bibnamefont {Senyshyn}}, \bibinfo {author}
  {\bibfnamefont {C.}~\bibnamefont {Baehtz}}, \ and\ \bibinfo {author}
  {\bibfnamefont {B.}~\bibnamefont {Albert}},\ }\href {\doibase
  https://doi.org/10.1002/zaac.200700376} {\bibfield  {journal} {\bibinfo
  {journal} {Zeitschrift für anorganische und allgemeine Chemie}\ }\textbf
  {\bibinfo {volume} {633}},\ \bibinfo {pages} {2626} (\bibinfo {year}
  {2007})}\BibitemShut {NoStop}%
\bibitem [{\citenamefont {Khlyustikov}\ and\ \citenamefont
  {Buzdin}(1988)}]{Buzdin1988}%
  \BibitemOpen
  \bibfield  {author} {\bibinfo {author} {\bibfnamefont {I.}~\bibnamefont
  {Khlyustikov}}\ and\ \bibinfo {author} {\bibfnamefont {A.}~\bibnamefont
  {Buzdin}},\ }\href {\doibase 10.1070/PU1988v031n05ABEH003545} {\bibfield
  {journal} {\bibinfo  {journal} {Soviet Physics Uspekhi}\ }\textbf {\bibinfo
  {volume} {31}},\ \bibinfo {pages} {409} (\bibinfo {year} {1988})}\BibitemShut
  {NoStop}%
\bibitem [{\citenamefont {Drozdov}\ \emph {et~al.}(2015)\citenamefont
  {Drozdov}, \citenamefont {Eremets}, \citenamefont {Troyan}, \citenamefont
  {Ksenofontov},\ and\ \citenamefont {Shylin}}]{h3sexp}%
  \BibitemOpen
  \bibfield  {author} {\bibinfo {author} {\bibfnamefont {A.~P.}\ \bibnamefont
  {Drozdov}}, \bibinfo {author} {\bibfnamefont {M.~I.}\ \bibnamefont
  {Eremets}}, \bibinfo {author} {\bibfnamefont {I.~A.}\ \bibnamefont {Troyan}},
  \bibinfo {author} {\bibfnamefont {V.}~\bibnamefont {Ksenofontov}}, \ and\
  \bibinfo {author} {\bibfnamefont {S.~I.}\ \bibnamefont {Shylin}},\ }\href
  {\doibase 10.1038/nature14964} {\bibfield  {journal} {\bibinfo  {journal}
  {Nature}\ }\textbf {\bibinfo {volume} {525}},\ \bibinfo {pages} {73}
  (\bibinfo {year} {2015})}\BibitemShut {NoStop}%
\bibitem [{\citenamefont {Somayazulu}\ \emph {et~al.}(2019)\citenamefont
  {Somayazulu}, \citenamefont {Ahart}, \citenamefont {Mishra}, \citenamefont
  {Geballe}, \citenamefont {Baldini}, \citenamefont {Meng}, \citenamefont
  {Struzhkin},\ and\ \citenamefont {Hemley}}]{Somayazulu}%
  \BibitemOpen
  \bibfield  {author} {\bibinfo {author} {\bibfnamefont {M.}~\bibnamefont
  {Somayazulu}}, \bibinfo {author} {\bibfnamefont {M.}~\bibnamefont {Ahart}},
  \bibinfo {author} {\bibfnamefont {A.~K.}\ \bibnamefont {Mishra}}, \bibinfo
  {author} {\bibfnamefont {Z.~M.}\ \bibnamefont {Geballe}}, \bibinfo {author}
  {\bibfnamefont {M.}~\bibnamefont {Baldini}}, \bibinfo {author} {\bibfnamefont
  {Y.}~\bibnamefont {Meng}}, \bibinfo {author} {\bibfnamefont {V.~V.}\
  \bibnamefont {Struzhkin}}, \ and\ \bibinfo {author} {\bibfnamefont {R.~J.}\
  \bibnamefont {Hemley}},\ }\href {\doibase 10.1103/PhysRevLett.122.027001}
  {\bibfield  {journal} {\bibinfo  {journal} {Phys. Rev. Lett.}\ }\textbf
  {\bibinfo {volume} {122}},\ \bibinfo {pages} {027001} (\bibinfo {year}
  {2019})}\BibitemShut {NoStop}%
\bibitem [{\citenamefont {Boeri}\ \emph {et~al.}(2021)\citenamefont {Boeri},
  \citenamefont {Hennig}, \citenamefont {Hirschfeld}, \citenamefont {Profeta},
  \citenamefont {Sanna}, \citenamefont {Zurek}, \citenamefont {Pickett},
  \citenamefont {Amsler}, \citenamefont {Dias}, \citenamefont {Eremets},
  \citenamefont {Heil}, \citenamefont {Hemley}, \citenamefont {Liu},
  \citenamefont {Ma}, \citenamefont {Pierleoni}, \citenamefont {Kolmogorov},
  \citenamefont {Rybin}, \citenamefont {Novoselov}, \citenamefont {Anisimov},
  \citenamefont {Oganov}, \citenamefont {Pickard}, \citenamefont {Bi},
  \citenamefont {Arita}, \citenamefont {Errea}, \citenamefont {Pellegrini},
  \citenamefont {Requist}, \citenamefont {Gross}, \citenamefont {Margine},
  \citenamefont {Xie}, \citenamefont {yundi quan}, \citenamefont {ajinkya
  hire}, \citenamefont {Fanfarillo}, \citenamefont {Stewart}, \citenamefont
  {Hamlin}, \citenamefont {Stanev}, \citenamefont {Gonnelli}, \citenamefont
  {Piatti}, \citenamefont {Romanin}, \citenamefont {Daghero},\ and\
  \citenamefont {Valenti}}]{Boeri2021}%
  \BibitemOpen
  \bibfield  {author} {\bibinfo {author} {\bibfnamefont {L.}~\bibnamefont
  {Boeri}}, \bibinfo {author} {\bibfnamefont {R.~G.}\ \bibnamefont {Hennig}},
  \bibinfo {author} {\bibfnamefont {P.~J.}\ \bibnamefont {Hirschfeld}},
  \bibinfo {author} {\bibfnamefont {G.}~\bibnamefont {Profeta}}, \bibinfo
  {author} {\bibfnamefont {A.}~\bibnamefont {Sanna}}, \bibinfo {author}
  {\bibfnamefont {E.}~\bibnamefont {Zurek}}, \bibinfo {author} {\bibfnamefont
  {W.~E.}\ \bibnamefont {Pickett}}, \bibinfo {author} {\bibfnamefont
  {M.}~\bibnamefont {Amsler}}, \bibinfo {author} {\bibfnamefont
  {R.}~\bibnamefont {Dias}}, \bibinfo {author} {\bibfnamefont {M.}~\bibnamefont
  {Eremets}}, \bibinfo {author} {\bibfnamefont {C.}~\bibnamefont {Heil}},
  \bibinfo {author} {\bibfnamefont {R.}~\bibnamefont {Hemley}}, \bibinfo
  {author} {\bibfnamefont {H.}~\bibnamefont {Liu}}, \bibinfo {author}
  {\bibfnamefont {Y.}~\bibnamefont {Ma}}, \bibinfo {author} {\bibfnamefont
  {C.}~\bibnamefont {Pierleoni}}, \bibinfo {author} {\bibfnamefont
  {A.}~\bibnamefont {Kolmogorov}}, \bibinfo {author} {\bibfnamefont
  {N.}~\bibnamefont {Rybin}}, \bibinfo {author} {\bibfnamefont
  {D.}~\bibnamefont {Novoselov}}, \bibinfo {author} {\bibfnamefont {V.~I.}\
  \bibnamefont {Anisimov}}, \bibinfo {author} {\bibfnamefont {A.~R.}\
  \bibnamefont {Oganov}}, \bibinfo {author} {\bibfnamefont {C.~J.}\
  \bibnamefont {Pickard}}, \bibinfo {author} {\bibfnamefont {T.}~\bibnamefont
  {Bi}}, \bibinfo {author} {\bibfnamefont {R.}~\bibnamefont {Arita}}, \bibinfo
  {author} {\bibfnamefont {I.}~\bibnamefont {Errea}}, \bibinfo {author}
  {\bibfnamefont {C.}~\bibnamefont {Pellegrini}}, \bibinfo {author}
  {\bibfnamefont {R.}~\bibnamefont {Requist}}, \bibinfo {author} {\bibfnamefont
  {E.}~\bibnamefont {Gross}}, \bibinfo {author} {\bibfnamefont {E.~R.}\
  \bibnamefont {Margine}}, \bibinfo {author} {\bibfnamefont {S.~R.}\
  \bibnamefont {Xie}}, \bibinfo {author} {\bibnamefont {yundi quan}}, \bibinfo
  {author} {\bibnamefont {ajinkya hire}}, \bibinfo {author} {\bibfnamefont
  {L.}~\bibnamefont {Fanfarillo}}, \bibinfo {author} {\bibfnamefont {G.~R.}\
  \bibnamefont {Stewart}}, \bibinfo {author} {\bibfnamefont {J.~J.}\
  \bibnamefont {Hamlin}}, \bibinfo {author} {\bibfnamefont {V.}~\bibnamefont
  {Stanev}}, \bibinfo {author} {\bibfnamefont {R.~S.}\ \bibnamefont
  {Gonnelli}}, \bibinfo {author} {\bibfnamefont {E.}~\bibnamefont {Piatti}},
  \bibinfo {author} {\bibfnamefont {D.}~\bibnamefont {Romanin}}, \bibinfo
  {author} {\bibfnamefont {D.}~\bibnamefont {Daghero}}, \ and\ \bibinfo
  {author} {\bibfnamefont {R.}~\bibnamefont {Valenti}},\ }\href {\doibase
  10.1088/1361-648x/ac2864} {\bibfield  {journal} {\bibinfo  {journal} {Journal
  of Physics: Condensed Matter}\ } (\bibinfo {year} {2021}),\
  10.1088/1361-648x/ac2864}\BibitemShut {NoStop}%
\bibitem [{\citenamefont {Vinet}\ \emph {et~al.}(1987)\citenamefont {Vinet},
  \citenamefont {Ferrante}, \citenamefont {Rose},\ and\ \citenamefont
  {Smith}}]{Vinet1987}%
  \BibitemOpen
  \bibfield  {author} {\bibinfo {author} {\bibfnamefont {P.}~\bibnamefont
  {Vinet}}, \bibinfo {author} {\bibfnamefont {J.}~\bibnamefont {Ferrante}},
  \bibinfo {author} {\bibfnamefont {J.~H.}\ \bibnamefont {Rose}}, \ and\
  \bibinfo {author} {\bibfnamefont {J.~R.}\ \bibnamefont {Smith}},\ }\href
  {\doibase https://doi.org/10.1029/JB092iB09p09319} {\bibfield  {journal}
  {\bibinfo  {journal} {Journal of Geophysical Research: Solid Earth}\ }\textbf
  {\bibinfo {volume} {92}},\ \bibinfo {pages} {9319} (\bibinfo {year}
  {1987})}\BibitemShut {NoStop}%
\bibitem [{\citenamefont {Perdew}\ \emph {et~al.}(1996)\citenamefont {Perdew},
  \citenamefont {Burke},\ and\ \citenamefont {Ernzerhof}}]{PBE}%
  \BibitemOpen
  \bibfield  {author} {\bibinfo {author} {\bibfnamefont {J.~P.}\ \bibnamefont
  {Perdew}}, \bibinfo {author} {\bibfnamefont {K.}~\bibnamefont {Burke}}, \
  and\ \bibinfo {author} {\bibfnamefont {M.}~\bibnamefont {Ernzerhof}},\ }\href
  {\doibase 10.1103/PhysRevLett.77.3865} {\bibfield  {journal} {\bibinfo
  {journal} {Phys. Rev. Lett.}\ }\textbf {\bibinfo {volume} {77}},\ \bibinfo
  {pages} {3865} (\bibinfo {year} {1996})}\BibitemShut {NoStop}%
\bibitem [{\citenamefont {Zhang}\ \emph {et~al.}(2021)\citenamefont {Zhang},
  \citenamefont {Xu}, \citenamefont {Li}, \citenamefont {Xu}, \citenamefont
  {Zhang}, \citenamefont {Greenberg}, \citenamefont {Prakapenka}, \citenamefont
  {Chen}, \citenamefont {Wuttig}, \citenamefont {Mao},\ and\ \citenamefont
  {Yang}}]{Zhang2017}%
  \BibitemOpen
  \bibfield  {author} {\bibinfo {author} {\bibfnamefont {K.}~\bibnamefont
  {Zhang}}, \bibinfo {author} {\bibfnamefont {M.}~\bibnamefont {Xu}}, \bibinfo
  {author} {\bibfnamefont {N.}~\bibnamefont {Li}}, \bibinfo {author}
  {\bibfnamefont {M.}~\bibnamefont {Xu}}, \bibinfo {author} {\bibfnamefont
  {Q.}~\bibnamefont {Zhang}}, \bibinfo {author} {\bibfnamefont
  {E.}~\bibnamefont {Greenberg}}, \bibinfo {author} {\bibfnamefont {V.~B.}\
  \bibnamefont {Prakapenka}}, \bibinfo {author} {\bibfnamefont {Y.-S.}\
  \bibnamefont {Chen}}, \bibinfo {author} {\bibfnamefont {M.}~\bibnamefont
  {Wuttig}}, \bibinfo {author} {\bibfnamefont {H.-K.}\ \bibnamefont {Mao}}, \
  and\ \bibinfo {author} {\bibfnamefont {W.}~\bibnamefont {Yang}},\ }\href
  {\doibase 10.1103/PhysRevLett.127.127002} {\bibfield  {journal} {\bibinfo
  {journal} {Phys. Rev. Lett.}\ }\textbf {\bibinfo {volume} {127}},\ \bibinfo
  {pages} {127002} (\bibinfo {year} {2021})}\BibitemShut {NoStop}%
\bibitem [{\citenamefont {Wang}\ \emph {et~al.}(2017)\citenamefont {Wang},
  \citenamefont {Tao}, \citenamefont {Ma}, \citenamefont {Cui}, \citenamefont
  {Wang}, \citenamefont {Dong},\ and\ \citenamefont
  {Zhu}}]{Wang_WB2ambient_2017}%
  \BibitemOpen
  \bibfield  {author} {\bibinfo {author} {\bibfnamefont {C.}~\bibnamefont
  {Wang}}, \bibinfo {author} {\bibfnamefont {Q.}~\bibnamefont {Tao}}, \bibinfo
  {author} {\bibfnamefont {S.}~\bibnamefont {Ma}}, \bibinfo {author}
  {\bibfnamefont {T.}~\bibnamefont {Cui}}, \bibinfo {author} {\bibfnamefont
  {X.}~\bibnamefont {Wang}}, \bibinfo {author} {\bibfnamefont {S.}~\bibnamefont
  {Dong}}, \ and\ \bibinfo {author} {\bibfnamefont {P.}~\bibnamefont {Zhu}},\
  }\href {\doibase 10.1039/C6CP04287B} {\bibfield  {journal} {\bibinfo
  {journal} {Phys. Chem. Chem. Phys.}\ }\textbf {\bibinfo {volume} {19}},\
  \bibinfo {pages} {8919} (\bibinfo {year} {2017})}\BibitemShut {NoStop}%
\bibitem [{\citenamefont {Hennig}\ \emph {et~al.}(2010)\citenamefont {Hennig},
  \citenamefont {Wadehra}, \citenamefont {Driver}, \citenamefont {Parker},
  \citenamefont {Umrigar},\ and\ \citenamefont {Wilkins}}]{PhysRevB.82.014101}%
  \BibitemOpen
  \bibfield  {author} {\bibinfo {author} {\bibfnamefont {R.~G.}\ \bibnamefont
  {Hennig}}, \bibinfo {author} {\bibfnamefont {A.}~\bibnamefont {Wadehra}},
  \bibinfo {author} {\bibfnamefont {K.~P.}\ \bibnamefont {Driver}}, \bibinfo
  {author} {\bibfnamefont {W.~D.}\ \bibnamefont {Parker}}, \bibinfo {author}
  {\bibfnamefont {C.~J.}\ \bibnamefont {Umrigar}}, \ and\ \bibinfo {author}
  {\bibfnamefont {J.~W.}\ \bibnamefont {Wilkins}},\ }\href {\doibase
  10.1103/PhysRevB.82.014101} {\bibfield  {journal} {\bibinfo  {journal} {Phys.
  Rev. B}\ }\textbf {\bibinfo {volume} {82}},\ \bibinfo {pages} {014101}
  (\bibinfo {year} {2010})}\BibitemShut {NoStop}%
\bibitem [{\citenamefont {Yin}\ \emph {et~al.}(2013)\citenamefont {Yin},
  \citenamefont {He}, \citenamefont {Xu}, \citenamefont {Wang}, \citenamefont
  {Wang}, \citenamefont {Li}, \citenamefont {Zhang}, \citenamefont {Liu},
  \citenamefont {Liu}, \citenamefont {Wang}, \citenamefont {Meng},\ and\
  \citenamefont {Zhu}}]{Yin_moduliWB2_2013}%
  \BibitemOpen
  \bibfield  {author} {\bibinfo {author} {\bibfnamefont {S.}~\bibnamefont
  {Yin}}, \bibinfo {author} {\bibfnamefont {D.}~\bibnamefont {He}}, \bibinfo
  {author} {\bibfnamefont {C.}~\bibnamefont {Xu}}, \bibinfo {author}
  {\bibfnamefont {W.}~\bibnamefont {Wang}}, \bibinfo {author} {\bibfnamefont
  {H.}~\bibnamefont {Wang}}, \bibinfo {author} {\bibfnamefont {L.}~\bibnamefont
  {Li}}, \bibinfo {author} {\bibfnamefont {L.}~\bibnamefont {Zhang}}, \bibinfo
  {author} {\bibfnamefont {F.}~\bibnamefont {Liu}}, \bibinfo {author}
  {\bibfnamefont {P.}~\bibnamefont {Liu}}, \bibinfo {author} {\bibfnamefont
  {Z.}~\bibnamefont {Wang}}, \bibinfo {author} {\bibfnamefont {C.}~\bibnamefont
  {Meng}}, \ and\ \bibinfo {author} {\bibfnamefont {W.}~\bibnamefont {Zhu}},\
  }\href {\doibase 10.1080/08957959.2013.791289} {\bibfield  {journal}
  {\bibinfo  {journal} {High Pressure Research}\ }\textbf {\bibinfo {volume}
  {33}},\ \bibinfo {pages} {409} (\bibinfo {year} {2013})},\ \Eprint
  {http://arxiv.org/abs/https://doi.org/10.1080/08957959.2013.791289}
  {https://doi.org/10.1080/08957959.2013.791289} \BibitemShut {NoStop}%
\bibitem [{\citenamefont {Zhang}\ \emph {et~al.}(2018)\citenamefont {Zhang},
  \citenamefont {Kitchaev}, \citenamefont {Yang}, \citenamefont {Chen},
  \citenamefont {Dacek}, \citenamefont {Sarmiento-P{\'{e}}rez}, \citenamefont
  {Marques}, \citenamefont {Peng}, \citenamefont {Ceder}, \citenamefont
  {Perdew},\ and\ \citenamefont {Sun}}]{Zhang2018}%
  \BibitemOpen
  \bibfield  {author} {\bibinfo {author} {\bibfnamefont {Y.}~\bibnamefont
  {Zhang}}, \bibinfo {author} {\bibfnamefont {D.~A.}\ \bibnamefont {Kitchaev}},
  \bibinfo {author} {\bibfnamefont {J.}~\bibnamefont {Yang}}, \bibinfo {author}
  {\bibfnamefont {T.}~\bibnamefont {Chen}}, \bibinfo {author} {\bibfnamefont
  {S.~T.}\ \bibnamefont {Dacek}}, \bibinfo {author} {\bibfnamefont {R.~A.}\
  \bibnamefont {Sarmiento-P{\'{e}}rez}}, \bibinfo {author} {\bibfnamefont
  {M.~A.~L.}\ \bibnamefont {Marques}}, \bibinfo {author} {\bibfnamefont
  {H.}~\bibnamefont {Peng}}, \bibinfo {author} {\bibfnamefont {G.}~\bibnamefont
  {Ceder}}, \bibinfo {author} {\bibfnamefont {J.~P.}\ \bibnamefont {Perdew}}, \
  and\ \bibinfo {author} {\bibfnamefont {J.}~\bibnamefont {Sun}},\ }\href
  {\doibase 10.1038/s41524-018-0065-z} {\bibfield  {journal} {\bibinfo
  {journal} {npj Computational Materials}\ }\textbf {\bibinfo {volume} {4}}
  (\bibinfo {year} {2018}),\ 10.1038/s41524-018-0065-z}\BibitemShut {NoStop}%
\bibitem [{\citenamefont {Yang}\ \emph {et~al.}(2019)\citenamefont {Yang},
  \citenamefont {Kitchaev},\ and\ \citenamefont {Ceder}}]{Yang2019}%
  \BibitemOpen
  \bibfield  {author} {\bibinfo {author} {\bibfnamefont {J.~H.}\ \bibnamefont
  {Yang}}, \bibinfo {author} {\bibfnamefont {D.~A.}\ \bibnamefont {Kitchaev}},
  \ and\ \bibinfo {author} {\bibfnamefont {G.}~\bibnamefont {Ceder}},\ }\href
  {\doibase 10.1103/physrevb.100.035132} {\bibfield  {journal} {\bibinfo
  {journal} {Physical Review B}\ }\textbf {\bibinfo {volume} {100}} (\bibinfo
  {year} {2019}),\ 10.1103/physrevb.100.035132}\BibitemShut {NoStop}%
\bibitem [{\citenamefont {Hinuma}\ \emph {et~al.}(2017)\citenamefont {Hinuma},
  \citenamefont {Hayashi}, \citenamefont {Kumagai}, \citenamefont {Tanaka},\
  and\ \citenamefont {Oba}}]{Hinuma2017}%
  \BibitemOpen
  \bibfield  {author} {\bibinfo {author} {\bibfnamefont {Y.}~\bibnamefont
  {Hinuma}}, \bibinfo {author} {\bibfnamefont {H.}~\bibnamefont {Hayashi}},
  \bibinfo {author} {\bibfnamefont {Y.}~\bibnamefont {Kumagai}}, \bibinfo
  {author} {\bibfnamefont {I.}~\bibnamefont {Tanaka}}, \ and\ \bibinfo {author}
  {\bibfnamefont {F.}~\bibnamefont {Oba}},\ }\href {\doibase
  10.1103/physrevb.96.094102} {\bibfield  {journal} {\bibinfo  {journal}
  {Physical Review B}\ }\textbf {\bibinfo {volume} {96}} (\bibinfo {year}
  {2017}),\ 10.1103/physrevb.96.094102}\BibitemShut {NoStop}%
\bibitem [{\citenamefont {Bartel}\ \emph {et~al.}(2019)\citenamefont {Bartel},
  \citenamefont {Weimer}, \citenamefont {Lany}, \citenamefont {Musgrave},\ and\
  \citenamefont {Holder}}]{Bartel2019}%
  \BibitemOpen
  \bibfield  {author} {\bibinfo {author} {\bibfnamefont {C.~J.}\ \bibnamefont
  {Bartel}}, \bibinfo {author} {\bibfnamefont {A.~W.}\ \bibnamefont {Weimer}},
  \bibinfo {author} {\bibfnamefont {S.}~\bibnamefont {Lany}}, \bibinfo {author}
  {\bibfnamefont {C.~B.}\ \bibnamefont {Musgrave}}, \ and\ \bibinfo {author}
  {\bibfnamefont {A.~M.}\ \bibnamefont {Holder}},\ }\href {\doibase
  10.1038/s41524-018-0143-2} {\bibfield  {journal} {\bibinfo  {journal} {npj
  Computational Materials}\ }\textbf {\bibinfo {volume} {5}} (\bibinfo {year}
  {2019}),\ 10.1038/s41524-018-0143-2}\BibitemShut {NoStop}%
\bibitem [{\citenamefont {{Xie}}\ \emph {et~al.}(2021)\citenamefont {{Xie}},
  \citenamefont {{Quan}}, \citenamefont {{Hire}}, \citenamefont {{Deng}},
  \citenamefont {{DeStefano}}, \citenamefont {{Salinas}}, \citenamefont
  {{Shah}}, \citenamefont {{Fanfarillo}}, \citenamefont {{Lim}}, \citenamefont
  {{Kim}}, \citenamefont {{Stewart}}, \citenamefont {{Hamlin}}, \citenamefont
  {{Hirschfeld}},\ and\ \citenamefont {{Hennig}}}]{Xie2021}%
  \BibitemOpen
  \bibfield  {author} {\bibinfo {author} {\bibfnamefont {S.~R.}\ \bibnamefont
  {{Xie}}}, \bibinfo {author} {\bibfnamefont {Y.}~\bibnamefont {{Quan}}},
  \bibinfo {author} {\bibfnamefont {A.~C.}\ \bibnamefont {{Hire}}}, \bibinfo
  {author} {\bibfnamefont {B.}~\bibnamefont {{Deng}}}, \bibinfo {author}
  {\bibfnamefont {J.~M.}\ \bibnamefont {{DeStefano}}}, \bibinfo {author}
  {\bibfnamefont {I.}~\bibnamefont {{Salinas}}}, \bibinfo {author}
  {\bibfnamefont {U.~S.}\ \bibnamefont {{Shah}}}, \bibinfo {author}
  {\bibfnamefont {L.}~\bibnamefont {{Fanfarillo}}}, \bibinfo {author}
  {\bibfnamefont {J.}~\bibnamefont {{Lim}}}, \bibinfo {author} {\bibfnamefont
  {J.}~\bibnamefont {{Kim}}}, \bibinfo {author} {\bibfnamefont {G.~R.}\
  \bibnamefont {{Stewart}}}, \bibinfo {author} {\bibfnamefont {J.~J.}\
  \bibnamefont {{Hamlin}}}, \bibinfo {author} {\bibfnamefont {P.~J.}\
  \bibnamefont {{Hirschfeld}}}, \ and\ \bibinfo {author} {\bibfnamefont
  {R.~G.}\ \bibnamefont {{Hennig}}},\ }\href@noop {} {\bibfield  {journal}
  {\bibinfo  {journal} {arXiv e-prints}\ ,\ \bibinfo {eid} {arXiv:2106.05235}}
  (\bibinfo {year} {2021})},\ \Eprint {http://arxiv.org/abs/2106.05235}
  {arXiv:2106.05235 [cond-mat.supr-con]} \BibitemShut {NoStop}%
\bibitem [{\citenamefont {Allen}\ and\ \citenamefont
  {Dynes}(1975)}]{Allen-Dynes1975}%
  \BibitemOpen
  \bibfield  {author} {\bibinfo {author} {\bibfnamefont {P.~B.}\ \bibnamefont
  {Allen}}\ and\ \bibinfo {author} {\bibfnamefont {R.~C.}\ \bibnamefont
  {Dynes}},\ }\href {\doibase 10.1103/PhysRevB.12.905} {\bibfield  {journal}
  {\bibinfo  {journal} {Phys. Rev. B}\ }\textbf {\bibinfo {volume} {12}},\
  \bibinfo {pages} {905} (\bibinfo {year} {1975})}\BibitemShut {NoStop}%
\bibitem [{\citenamefont {Mazin}\ and\ \citenamefont
  {Antropov}(2003)}]{Mazin2003}%
  \BibitemOpen
  \bibfield  {author} {\bibinfo {author} {\bibfnamefont {I.}~\bibnamefont
  {Mazin}}\ and\ \bibinfo {author} {\bibfnamefont {V.}~\bibnamefont
  {Antropov}},\ }\href {\doibase https://doi.org/10.1016/S0921-4534(02)02299-2}
  {\bibfield  {journal} {\bibinfo  {journal} {Physica C: Superconductivity}\
  }\textbf {\bibinfo {volume} {385}},\ \bibinfo {pages} {49} (\bibinfo {year}
  {2003})}\BibitemShut {NoStop}%
\bibitem [{\citenamefont {Heuser}\ \emph {et~al.}(2000)\citenamefont {Heuser},
  \citenamefont {Scheidt}, \citenamefont {Schreiner}, \citenamefont {Fisk},\
  and\ \citenamefont {Stewart}}]{Heuser2000}%
  \BibitemOpen
  \bibfield  {author} {\bibinfo {author} {\bibfnamefont {K.}~\bibnamefont
  {Heuser}}, \bibinfo {author} {\bibfnamefont {E.-W.}\ \bibnamefont {Scheidt}},
  \bibinfo {author} {\bibfnamefont {T.}~\bibnamefont {Schreiner}}, \bibinfo
  {author} {\bibfnamefont {Z.}~\bibnamefont {Fisk}}, \ and\ \bibinfo {author}
  {\bibfnamefont {G.}~\bibnamefont {Stewart}},\ }\href {\doibase
  10.1023/A:1004651225026} {\bibfield  {journal} {\bibinfo  {journal} {Journal
  of Low Temperature Physics}\ }\textbf {\bibinfo {volume} {118}},\ \bibinfo
  {pages} {235} (\bibinfo {year} {2000})}\BibitemShut {NoStop}%
\bibitem [{\citenamefont {Prescher}\ and\ \citenamefont
  {Prakapenka}(2015)}]{Dioptas2015}%
  \BibitemOpen
  \bibfield  {author} {\bibinfo {author} {\bibfnamefont {C.}~\bibnamefont
  {Prescher}}\ and\ \bibinfo {author} {\bibfnamefont {V.~B.}\ \bibnamefont
  {Prakapenka}},\ }\href {https://doi.org/10.1080/08957959.2015.1059835}
  {\bibfield  {journal} {\bibinfo  {journal} {High Pressure Research}\ }\textbf
  {\bibinfo {volume} {35}},\ \bibinfo {pages} {223} (\bibinfo {year}
  {2015})}\BibitemShut {NoStop}%
\bibitem [{\citenamefont {Rietveld}(1969)}]{Rietveld_1969}%
  \BibitemOpen
  \bibfield  {author} {\bibinfo {author} {\bibfnamefont {H.~M.}\ \bibnamefont
  {Rietveld}},\ }\href {\doibase 10.1107/S0021889869006558} {\bibfield
  {journal} {\bibinfo  {journal} {Journal of Applied Crystallography}\ }\textbf
  {\bibinfo {volume} {2}},\ \bibinfo {pages} {65} (\bibinfo {year}
  {1969})}\BibitemShut {NoStop}%
\bibitem [{\citenamefont {{Le Bail}}\ \emph {et~al.}(1988)\citenamefont {{Le
  Bail}}, \citenamefont {Duroy},\ and\ \citenamefont {Fourquet}}]{LEBAIL1988}%
  \BibitemOpen
  \bibfield  {author} {\bibinfo {author} {\bibfnamefont {A.}~\bibnamefont {{Le
  Bail}}}, \bibinfo {author} {\bibfnamefont {H.}~\bibnamefont {Duroy}}, \ and\
  \bibinfo {author} {\bibfnamefont {J.}~\bibnamefont {Fourquet}},\ }\href
  {\doibase https://doi.org/10.1016/0025-5408(88)90019-0} {\bibfield  {journal}
  {\bibinfo  {journal} {Materials Research Bulletin}\ }\textbf {\bibinfo
  {volume} {23}},\ \bibinfo {pages} {447 } (\bibinfo {year}
  {1988})}\BibitemShut {NoStop}%
\bibitem [{\citenamefont {Toby}\ and\ \citenamefont
  {Von~Dreele}(2013)}]{Toby_GSASII_2013}%
  \BibitemOpen
  \bibfield  {author} {\bibinfo {author} {\bibfnamefont {B.~H.}\ \bibnamefont
  {Toby}}\ and\ \bibinfo {author} {\bibfnamefont {R.~B.}\ \bibnamefont
  {Von~Dreele}},\ }\href {\doibase 10.1107/S0021889813003531} {\bibfield
  {journal} {\bibinfo  {journal} {Journal of Applied Crystallography}\ }\textbf
  {\bibinfo {volume} {46}},\ \bibinfo {pages} {544} (\bibinfo {year}
  {2013})}\BibitemShut {NoStop}%
\bibitem [{\citenamefont {Momma}\ and\ \citenamefont
  {Izumi}(2008)}]{Momma_VESTA_2008}%
  \BibitemOpen
  \bibfield  {author} {\bibinfo {author} {\bibfnamefont {K.}~\bibnamefont
  {Momma}}\ and\ \bibinfo {author} {\bibfnamefont {F.}~\bibnamefont {Izumi}},\
  }\href {\doibase 10.1107/S0021889808012016} {\bibfield  {journal} {\bibinfo
  {journal} {Journal of Applied Crystallography}\ }\textbf {\bibinfo {volume}
  {41}},\ \bibinfo {pages} {653} (\bibinfo {year} {2008})}\BibitemShut
  {NoStop}%
\bibitem [{\citenamefont {Chijioke}\ \emph {et~al.}(2005)\citenamefont
  {Chijioke}, \citenamefont {Nellis}, \citenamefont {Soldatov},\ and\
  \citenamefont {Silvera}}]{chijioke_ruby_2005}%
  \BibitemOpen
  \bibfield  {author} {\bibinfo {author} {\bibfnamefont {A.~D.}\ \bibnamefont
  {Chijioke}}, \bibinfo {author} {\bibfnamefont {W.~J.}\ \bibnamefont
  {Nellis}}, \bibinfo {author} {\bibfnamefont {A.}~\bibnamefont {Soldatov}}, \
  and\ \bibinfo {author} {\bibfnamefont {I.~F.}\ \bibnamefont {Silvera}},\
  }\href {\doibase 10.1063/1.2135877} {\bibfield  {journal} {\bibinfo
  {journal} {Journal of Applied Physics}\ }\textbf {\bibinfo {volume} {98}},\
  \bibinfo {pages} {114905} (\bibinfo {year} {2005})}\BibitemShut {NoStop}%
\bibitem [{\citenamefont {Akahama}\ and\ \citenamefont
  {Kawamura}(2006)}]{Akahama2006}%
  \BibitemOpen
  \bibfield  {author} {\bibinfo {author} {\bibfnamefont {Y.}~\bibnamefont
  {Akahama}}\ and\ \bibinfo {author} {\bibfnamefont {H.}~\bibnamefont
  {Kawamura}},\ }\href {\doibase 10.1063/1.2335683} {\bibfield  {journal}
  {\bibinfo  {journal} {Journal of Applied Physics}\ }\textbf {\bibinfo
  {volume} {100}},\ \bibinfo {pages} {043516} (\bibinfo {year} {2006})},\
  \Eprint {http://arxiv.org/abs/https://doi.org/10.1063/1.2335683}
  {https://doi.org/10.1063/1.2335683} \BibitemShut {NoStop}%
\bibitem [{\citenamefont {Matsuoka}\ and\ \citenamefont
  {Shimizu}(2009)}]{Matsuoka2009}%
  \BibitemOpen
  \bibfield  {author} {\bibinfo {author} {\bibfnamefont {T.}~\bibnamefont
  {Matsuoka}}\ and\ \bibinfo {author} {\bibfnamefont {K.}~\bibnamefont
  {Shimizu}},\ }\href {\doibase 10.1038/nature07827} {\bibfield  {journal}
  {\bibinfo  {journal} {Nature}\ }\textbf {\bibinfo {volume} {458}},\ \bibinfo
  {pages} {186} (\bibinfo {year} {2009})}\BibitemShut {NoStop}%
\bibitem [{\citenamefont {Weir}\ \emph {et~al.}(2000)\citenamefont {Weir},
  \citenamefont {Akella}, \citenamefont {Aracne-Ruddle}, \citenamefont
  {Vohra},\ and\ \citenamefont {Catledge}}]{weir_epitaxial_2000}%
  \BibitemOpen
  \bibfield  {author} {\bibinfo {author} {\bibfnamefont {S.~T.}\ \bibnamefont
  {Weir}}, \bibinfo {author} {\bibfnamefont {J.}~\bibnamefont {Akella}},
  \bibinfo {author} {\bibfnamefont {C.}~\bibnamefont {Aracne-Ruddle}}, \bibinfo
  {author} {\bibfnamefont {Y.~K.}\ \bibnamefont {Vohra}}, \ and\ \bibinfo
  {author} {\bibfnamefont {S.~A.}\ \bibnamefont {Catledge}},\ }\href {\doibase
  10.1063/1.1326838} {\bibfield  {journal} {\bibinfo  {journal} {Applied
  Physics Letters}\ }\textbf {\bibinfo {volume} {77}},\ \bibinfo {pages} {3400}
  (\bibinfo {year} {2000})}\BibitemShut {NoStop}%
\bibitem [{\citenamefont {Kresse}\ and\ \citenamefont
  {Furthm\"{u}ller}(1996{\natexlab{a}})}]{Kresse1996}%
  \BibitemOpen
  \bibfield  {author} {\bibinfo {author} {\bibfnamefont {G.}~\bibnamefont
  {Kresse}}\ and\ \bibinfo {author} {\bibfnamefont {J.}~\bibnamefont
  {Furthm\"{u}ller}},\ }\href {\doibase 10.1016/0927-0256(96)00008-0}
  {\bibfield  {journal} {\bibinfo  {journal} {Computational Materials Science}\
  }\textbf {\bibinfo {volume} {6}},\ \bibinfo {pages} {15} (\bibinfo {year}
  {1996}{\natexlab{a}})}\BibitemShut {NoStop}%
\bibitem [{\citenamefont {Kresse}\ and\ \citenamefont
  {Furthm\"{u}ller}(1996{\natexlab{b}})}]{Kresse1996_2}%
  \BibitemOpen
  \bibfield  {author} {\bibinfo {author} {\bibfnamefont {G.}~\bibnamefont
  {Kresse}}\ and\ \bibinfo {author} {\bibfnamefont {J.}~\bibnamefont
  {Furthm\"{u}ller}},\ }\href {\doibase 10.1103/physrevb.54.11169} {\bibfield
  {journal} {\bibinfo  {journal} {Physical Review B}\ }\textbf {\bibinfo
  {volume} {54}},\ \bibinfo {pages} {11169} (\bibinfo {year}
  {1996}{\natexlab{b}})}\BibitemShut {NoStop}%
\bibitem [{\citenamefont {Csonka}\ \emph {et~al.}(2009)\citenamefont {Csonka},
  \citenamefont {Perdew}, \citenamefont {Ruzsinszky}, \citenamefont
  {Philipsen}, \citenamefont {Leb{\`{e}}gue}, \citenamefont {Paier},
  \citenamefont {Vydrov},\ and\ \citenamefont
  {{\'{A}}ngy{\'{a}}n}}]{Csonka2009}%
  \BibitemOpen
  \bibfield  {author} {\bibinfo {author} {\bibfnamefont {G.~I.}\ \bibnamefont
  {Csonka}}, \bibinfo {author} {\bibfnamefont {J.~P.}\ \bibnamefont {Perdew}},
  \bibinfo {author} {\bibfnamefont {A.}~\bibnamefont {Ruzsinszky}}, \bibinfo
  {author} {\bibfnamefont {P.~H.~T.}\ \bibnamefont {Philipsen}}, \bibinfo
  {author} {\bibfnamefont {S.}~\bibnamefont {Leb{\`{e}}gue}}, \bibinfo {author}
  {\bibfnamefont {J.}~\bibnamefont {Paier}}, \bibinfo {author} {\bibfnamefont
  {O.~A.}\ \bibnamefont {Vydrov}}, \ and\ \bibinfo {author} {\bibfnamefont
  {J.~G.}\ \bibnamefont {{\'{A}}ngy{\'{a}}n}},\ }\href {\doibase
  10.1103/physrevb.79.155107} {\bibfield  {journal} {\bibinfo  {journal}
  {Physical Review B}\ }\textbf {\bibinfo {volume} {79}} (\bibinfo {year}
  {2009}),\ 10.1103/physrevb.79.155107}\BibitemShut {NoStop}%
\bibitem [{\citenamefont {Perdew}\ \emph {et~al.}(2008)\citenamefont {Perdew},
  \citenamefont {Ruzsinszky}, \citenamefont {Csonka}, \citenamefont {Vydrov},
  \citenamefont {Scuseria}, \citenamefont {Constantin}, \citenamefont {Zhou},\
  and\ \citenamefont {Burke}}]{Perdew2008}%
  \BibitemOpen
  \bibfield  {author} {\bibinfo {author} {\bibfnamefont {J.~P.}\ \bibnamefont
  {Perdew}}, \bibinfo {author} {\bibfnamefont {A.}~\bibnamefont {Ruzsinszky}},
  \bibinfo {author} {\bibfnamefont {G.~I.}\ \bibnamefont {Csonka}}, \bibinfo
  {author} {\bibfnamefont {O.~A.}\ \bibnamefont {Vydrov}}, \bibinfo {author}
  {\bibfnamefont {G.~E.}\ \bibnamefont {Scuseria}}, \bibinfo {author}
  {\bibfnamefont {L.~A.}\ \bibnamefont {Constantin}}, \bibinfo {author}
  {\bibfnamefont {X.}~\bibnamefont {Zhou}}, \ and\ \bibinfo {author}
  {\bibfnamefont {K.}~\bibnamefont {Burke}},\ }\href {\doibase
  10.1103/physrevlett.100.136406} {\bibfield  {journal} {\bibinfo  {journal}
  {Physical Review Letters}\ }\textbf {\bibinfo {volume} {100}} (\bibinfo
  {year} {2008}),\ 10.1103/physrevlett.100.136406}\BibitemShut {NoStop}%
\bibitem [{\citenamefont {Bl\"{o}chl}(1994)}]{Blchl1994}%
  \BibitemOpen
  \bibfield  {author} {\bibinfo {author} {\bibfnamefont {P.~E.}\ \bibnamefont
  {Bl\"{o}chl}},\ }\href {\doibase 10.1103/physrevb.50.17953} {\bibfield
  {journal} {\bibinfo  {journal} {Physical Review B}\ }\textbf {\bibinfo
  {volume} {50}},\ \bibinfo {pages} {17953} (\bibinfo {year}
  {1994})}\BibitemShut {NoStop}%
\bibitem [{\citenamefont {Methfessel}\ and\ \citenamefont
  {Paxton}(1989)}]{Methfessel1989}%
  \BibitemOpen
  \bibfield  {author} {\bibinfo {author} {\bibfnamefont {M.}~\bibnamefont
  {Methfessel}}\ and\ \bibinfo {author} {\bibfnamefont {A.~T.}\ \bibnamefont
  {Paxton}},\ }\href {\doibase 10.1103/physrevb.40.3616} {\bibfield  {journal}
  {\bibinfo  {journal} {Physical Review B}\ }\textbf {\bibinfo {volume} {40}},\
  \bibinfo {pages} {3616} (\bibinfo {year} {1989})}\BibitemShut {NoStop}%
\bibitem [{\citenamefont {Bl\"{o}chl}\ \emph {et~al.}(1994)\citenamefont
  {Bl\"{o}chl}, \citenamefont {Jepsen},\ and\ \citenamefont
  {Andersen}}]{Blchl1994_paw}%
  \BibitemOpen
  \bibfield  {author} {\bibinfo {author} {\bibfnamefont {P.~E.}\ \bibnamefont
  {Bl\"{o}chl}}, \bibinfo {author} {\bibfnamefont {O.}~\bibnamefont {Jepsen}},
  \ and\ \bibinfo {author} {\bibfnamefont {O.~K.}\ \bibnamefont {Andersen}},\
  }\href {\doibase 10.1103/physrevb.49.16223} {\bibfield  {journal} {\bibinfo
  {journal} {Physical Review B}\ }\textbf {\bibinfo {volume} {49}},\ \bibinfo
  {pages} {16223} (\bibinfo {year} {1994})}\BibitemShut {NoStop}%
\bibitem [{\citenamefont {Giannozzi}\ \emph {et~al.}(2009)\citenamefont
  {Giannozzi}, \citenamefont {Baroni}, \citenamefont {Bonini}, \citenamefont
  {Calandra}, \citenamefont {Car}, \citenamefont {Cavazzoni}, \citenamefont
  {Ceresoli}, \citenamefont {Chiarotti}, \citenamefont {Cococcioni},
  \citenamefont {Dabo}, \citenamefont {Corso}, \citenamefont {de~Gironcoli},
  \citenamefont {Fabris}, \citenamefont {Fratesi}, \citenamefont {Gebauer},
  \citenamefont {Gerstmann}, \citenamefont {Gougoussis}, \citenamefont
  {Kokalj}, \citenamefont {Lazzeri}, \citenamefont {Martin-Samos},
  \citenamefont {Marzari}, \citenamefont {Mauri}, \citenamefont {Mazzarello},
  \citenamefont {Paolini}, \citenamefont {Pasquarello}, \citenamefont
  {Paulatto}, \citenamefont {Sbraccia}, \citenamefont {Scandolo}, \citenamefont
  {Sclauzero}, \citenamefont {Seitsonen}, \citenamefont {Smogunov},
  \citenamefont {Umari},\ and\ \citenamefont {Wentzcovitch}}]{qe1}%
  \BibitemOpen
  \bibfield  {author} {\bibinfo {author} {\bibfnamefont {P.}~\bibnamefont
  {Giannozzi}}, \bibinfo {author} {\bibfnamefont {S.}~\bibnamefont {Baroni}},
  \bibinfo {author} {\bibfnamefont {N.}~\bibnamefont {Bonini}}, \bibinfo
  {author} {\bibfnamefont {M.}~\bibnamefont {Calandra}}, \bibinfo {author}
  {\bibfnamefont {R.}~\bibnamefont {Car}}, \bibinfo {author} {\bibfnamefont
  {C.}~\bibnamefont {Cavazzoni}}, \bibinfo {author} {\bibfnamefont
  {D.}~\bibnamefont {Ceresoli}}, \bibinfo {author} {\bibfnamefont {G.~L.}\
  \bibnamefont {Chiarotti}}, \bibinfo {author} {\bibfnamefont {M.}~\bibnamefont
  {Cococcioni}}, \bibinfo {author} {\bibfnamefont {I.}~\bibnamefont {Dabo}},
  \bibinfo {author} {\bibfnamefont {A.~D.}\ \bibnamefont {Corso}}, \bibinfo
  {author} {\bibfnamefont {S.}~\bibnamefont {de~Gironcoli}}, \bibinfo {author}
  {\bibfnamefont {S.}~\bibnamefont {Fabris}}, \bibinfo {author} {\bibfnamefont
  {G.}~\bibnamefont {Fratesi}}, \bibinfo {author} {\bibfnamefont
  {R.}~\bibnamefont {Gebauer}}, \bibinfo {author} {\bibfnamefont
  {U.}~\bibnamefont {Gerstmann}}, \bibinfo {author} {\bibfnamefont
  {C.}~\bibnamefont {Gougoussis}}, \bibinfo {author} {\bibfnamefont
  {A.}~\bibnamefont {Kokalj}}, \bibinfo {author} {\bibfnamefont
  {M.}~\bibnamefont {Lazzeri}}, \bibinfo {author} {\bibfnamefont
  {L.}~\bibnamefont {Martin-Samos}}, \bibinfo {author} {\bibfnamefont
  {N.}~\bibnamefont {Marzari}}, \bibinfo {author} {\bibfnamefont
  {F.}~\bibnamefont {Mauri}}, \bibinfo {author} {\bibfnamefont
  {R.}~\bibnamefont {Mazzarello}}, \bibinfo {author} {\bibfnamefont
  {S.}~\bibnamefont {Paolini}}, \bibinfo {author} {\bibfnamefont
  {A.}~\bibnamefont {Pasquarello}}, \bibinfo {author} {\bibfnamefont
  {L.}~\bibnamefont {Paulatto}}, \bibinfo {author} {\bibfnamefont
  {C.}~\bibnamefont {Sbraccia}}, \bibinfo {author} {\bibfnamefont
  {S.}~\bibnamefont {Scandolo}}, \bibinfo {author} {\bibfnamefont
  {G.}~\bibnamefont {Sclauzero}}, \bibinfo {author} {\bibfnamefont {A.~P.}\
  \bibnamefont {Seitsonen}}, \bibinfo {author} {\bibfnamefont {A.}~\bibnamefont
  {Smogunov}}, \bibinfo {author} {\bibfnamefont {P.}~\bibnamefont {Umari}}, \
  and\ \bibinfo {author} {\bibfnamefont {R.~M.}\ \bibnamefont {Wentzcovitch}},\
  }\href {\doibase 10.1088/0953-8984/21/39/395502} {\bibfield  {journal}
  {\bibinfo  {journal} {Journal of Physics: Condensed Matter}\ }\textbf
  {\bibinfo {volume} {21}},\ \bibinfo {pages} {395502} (\bibinfo {year}
  {2009})}\BibitemShut {NoStop}%
\bibitem [{\citenamefont {Giannozzi}\ \emph {et~al.}(2020)\citenamefont
  {Giannozzi}, \citenamefont {Baseggio}, \citenamefont {Bonfà}, \citenamefont
  {Brunato}, \citenamefont {Car}, \citenamefont {Carnimeo}, \citenamefont
  {Cavazzoni}, \citenamefont {de~Gironcoli}, \citenamefont {Delugas},
  \citenamefont {Ferrari~Ruffino}, \citenamefont {Ferretti}, \citenamefont
  {Marzari}, \citenamefont {Timrov}, \citenamefont {Urru},\ and\ \citenamefont
  {Baroni}}]{qe2}%
  \BibitemOpen
  \bibfield  {author} {\bibinfo {author} {\bibfnamefont {P.}~\bibnamefont
  {Giannozzi}}, \bibinfo {author} {\bibfnamefont {O.}~\bibnamefont {Baseggio}},
  \bibinfo {author} {\bibfnamefont {P.}~\bibnamefont {Bonfà}}, \bibinfo
  {author} {\bibfnamefont {D.}~\bibnamefont {Brunato}}, \bibinfo {author}
  {\bibfnamefont {R.}~\bibnamefont {Car}}, \bibinfo {author} {\bibfnamefont
  {I.}~\bibnamefont {Carnimeo}}, \bibinfo {author} {\bibfnamefont
  {C.}~\bibnamefont {Cavazzoni}}, \bibinfo {author} {\bibfnamefont
  {S.}~\bibnamefont {de~Gironcoli}}, \bibinfo {author} {\bibfnamefont
  {P.}~\bibnamefont {Delugas}}, \bibinfo {author} {\bibfnamefont
  {F.}~\bibnamefont {Ferrari~Ruffino}}, \bibinfo {author} {\bibfnamefont
  {A.}~\bibnamefont {Ferretti}}, \bibinfo {author} {\bibfnamefont
  {N.}~\bibnamefont {Marzari}}, \bibinfo {author} {\bibfnamefont
  {I.}~\bibnamefont {Timrov}}, \bibinfo {author} {\bibfnamefont
  {A.}~\bibnamefont {Urru}}, \ and\ \bibinfo {author} {\bibfnamefont
  {S.}~\bibnamefont {Baroni}},\ }\href {\doibase 10.1063/5.0005082} {\bibfield
  {journal} {\bibinfo  {journal} {The Journal of Chemical Physics}\ }\textbf
  {\bibinfo {volume} {152}},\ \bibinfo {pages} {154105} (\bibinfo {year}
  {2020})},\ \Eprint {http://arxiv.org/abs/https://doi.org/10.1063/5.0005082}
  {https://doi.org/10.1063/5.0005082} \BibitemShut {NoStop}%
\bibitem [{\citenamefont {Giannozzi}\ \emph {et~al.}(2017)\citenamefont
  {Giannozzi}, \citenamefont {Andreussi}, \citenamefont {Brumme}, \citenamefont
  {Bunau}, \citenamefont {Nardelli}, \citenamefont {Calandra}, \citenamefont
  {Car}, \citenamefont {Cavazzoni}, \citenamefont {Ceresoli}, \citenamefont
  {Cococcioni}, \citenamefont {Colonna}, \citenamefont {Carnimeo},
  \citenamefont {Corso}, \citenamefont {de~Gironcoli}, \citenamefont {Delugas},
  \citenamefont {DiStasio}, \citenamefont {Ferretti}, \citenamefont {Floris},
  \citenamefont {Fratesi}, \citenamefont {Fugallo}, \citenamefont {Gebauer},
  \citenamefont {Gerstmann}, \citenamefont {Giustino}, \citenamefont {Gorni},
  \citenamefont {Jia}, \citenamefont {Kawamura}, \citenamefont {Ko},
  \citenamefont {Kokalj}, \citenamefont {Kü{\c{c}}ükbenli}, \citenamefont
  {Lazzeri}, \citenamefont {Marsili}, \citenamefont {Marzari}, \citenamefont
  {Mauri}, \citenamefont {Nguyen}, \citenamefont {Nguyen}, \citenamefont {de-la
  Roza}, \citenamefont {Paulatto}, \citenamefont {Ponc{\'{e}}}, \citenamefont
  {Rocca}, \citenamefont {Sabatini}, \citenamefont {Santra}, \citenamefont
  {Schlipf}, \citenamefont {Seitsonen}, \citenamefont {Smogunov}, \citenamefont
  {Timrov}, \citenamefont {Thonhauser}, \citenamefont {Umari}, \citenamefont
  {Vast}, \citenamefont {Wu},\ and\ \citenamefont {Baroni}}]{qe3}%
  \BibitemOpen
  \bibfield  {author} {\bibinfo {author} {\bibfnamefont {P.}~\bibnamefont
  {Giannozzi}}, \bibinfo {author} {\bibfnamefont {O.}~\bibnamefont
  {Andreussi}}, \bibinfo {author} {\bibfnamefont {T.}~\bibnamefont {Brumme}},
  \bibinfo {author} {\bibfnamefont {O.}~\bibnamefont {Bunau}}, \bibinfo
  {author} {\bibfnamefont {M.~B.}\ \bibnamefont {Nardelli}}, \bibinfo {author}
  {\bibfnamefont {M.}~\bibnamefont {Calandra}}, \bibinfo {author}
  {\bibfnamefont {R.}~\bibnamefont {Car}}, \bibinfo {author} {\bibfnamefont
  {C.}~\bibnamefont {Cavazzoni}}, \bibinfo {author} {\bibfnamefont
  {D.}~\bibnamefont {Ceresoli}}, \bibinfo {author} {\bibfnamefont
  {M.}~\bibnamefont {Cococcioni}}, \bibinfo {author} {\bibfnamefont
  {N.}~\bibnamefont {Colonna}}, \bibinfo {author} {\bibfnamefont
  {I.}~\bibnamefont {Carnimeo}}, \bibinfo {author} {\bibfnamefont {A.~D.}\
  \bibnamefont {Corso}}, \bibinfo {author} {\bibfnamefont {S.}~\bibnamefont
  {de~Gironcoli}}, \bibinfo {author} {\bibfnamefont {P.}~\bibnamefont
  {Delugas}}, \bibinfo {author} {\bibfnamefont {R.~A.}\ \bibnamefont
  {DiStasio}}, \bibinfo {author} {\bibfnamefont {A.}~\bibnamefont {Ferretti}},
  \bibinfo {author} {\bibfnamefont {A.}~\bibnamefont {Floris}}, \bibinfo
  {author} {\bibfnamefont {G.}~\bibnamefont {Fratesi}}, \bibinfo {author}
  {\bibfnamefont {G.}~\bibnamefont {Fugallo}}, \bibinfo {author} {\bibfnamefont
  {R.}~\bibnamefont {Gebauer}}, \bibinfo {author} {\bibfnamefont
  {U.}~\bibnamefont {Gerstmann}}, \bibinfo {author} {\bibfnamefont
  {F.}~\bibnamefont {Giustino}}, \bibinfo {author} {\bibfnamefont
  {T.}~\bibnamefont {Gorni}}, \bibinfo {author} {\bibfnamefont
  {J.}~\bibnamefont {Jia}}, \bibinfo {author} {\bibfnamefont {M.}~\bibnamefont
  {Kawamura}}, \bibinfo {author} {\bibfnamefont {H.-Y.}\ \bibnamefont {Ko}},
  \bibinfo {author} {\bibfnamefont {A.}~\bibnamefont {Kokalj}}, \bibinfo
  {author} {\bibfnamefont {E.}~\bibnamefont {Kü{\c{c}}ükbenli}}, \bibinfo
  {author} {\bibfnamefont {M.}~\bibnamefont {Lazzeri}}, \bibinfo {author}
  {\bibfnamefont {M.}~\bibnamefont {Marsili}}, \bibinfo {author} {\bibfnamefont
  {N.}~\bibnamefont {Marzari}}, \bibinfo {author} {\bibfnamefont
  {F.}~\bibnamefont {Mauri}}, \bibinfo {author} {\bibfnamefont {N.~L.}\
  \bibnamefont {Nguyen}}, \bibinfo {author} {\bibfnamefont {H.-V.}\
  \bibnamefont {Nguyen}}, \bibinfo {author} {\bibfnamefont {A.~O.}\
  \bibnamefont {de-la Roza}}, \bibinfo {author} {\bibfnamefont
  {L.}~\bibnamefont {Paulatto}}, \bibinfo {author} {\bibfnamefont
  {S.}~\bibnamefont {Ponc{\'{e}}}}, \bibinfo {author} {\bibfnamefont
  {D.}~\bibnamefont {Rocca}}, \bibinfo {author} {\bibfnamefont
  {R.}~\bibnamefont {Sabatini}}, \bibinfo {author} {\bibfnamefont
  {B.}~\bibnamefont {Santra}}, \bibinfo {author} {\bibfnamefont
  {M.}~\bibnamefont {Schlipf}}, \bibinfo {author} {\bibfnamefont {A.~P.}\
  \bibnamefont {Seitsonen}}, \bibinfo {author} {\bibfnamefont {A.}~\bibnamefont
  {Smogunov}}, \bibinfo {author} {\bibfnamefont {I.}~\bibnamefont {Timrov}},
  \bibinfo {author} {\bibfnamefont {T.}~\bibnamefont {Thonhauser}}, \bibinfo
  {author} {\bibfnamefont {P.}~\bibnamefont {Umari}}, \bibinfo {author}
  {\bibfnamefont {N.}~\bibnamefont {Vast}}, \bibinfo {author} {\bibfnamefont
  {X.}~\bibnamefont {Wu}}, \ and\ \bibinfo {author} {\bibfnamefont
  {S.}~\bibnamefont {Baroni}},\ }\href {\doibase 10.1088/1361-648x/aa8f79}
  {\bibfield  {journal} {\bibinfo  {journal} {Journal of Physics: Condensed
  Matter}\ }\textbf {\bibinfo {volume} {29}},\ \bibinfo {pages} {465901}
  (\bibinfo {year} {2017})}\BibitemShut {NoStop}%
\bibitem [{\citenamefont {Hamann}(2013)}]{Hamann2013}%
  \BibitemOpen
  \bibfield  {author} {\bibinfo {author} {\bibfnamefont {D.~R.}\ \bibnamefont
  {Hamann}},\ }\href {\doibase 10.1103/physrevb.88.085117} {\bibfield
  {journal} {\bibinfo  {journal} {Physical Review B}\ }\textbf {\bibinfo
  {volume} {88}} (\bibinfo {year} {2013}),\
  10.1103/physrevb.88.085117}\BibitemShut {NoStop}%
\bibitem [{\citenamefont {Schlipf}\ and\ \citenamefont
  {Gygi}(2015)}]{Schlipf2015}%
  \BibitemOpen
  \bibfield  {author} {\bibinfo {author} {\bibfnamefont {M.}~\bibnamefont
  {Schlipf}}\ and\ \bibinfo {author} {\bibfnamefont {F.}~\bibnamefont {Gygi}},\
  }\href {\doibase 10.1016/j.cpc.2015.05.011} {\bibfield  {journal} {\bibinfo
  {journal} {Computer Physics Communications}\ }\textbf {\bibinfo {volume}
  {196}},\ \bibinfo {pages} {36} (\bibinfo {year} {2015})}\BibitemShut
  {NoStop}%
\bibitem [{\citenamefont {Kawamura}\ \emph {et~al.}(2014)\citenamefont
  {Kawamura}, \citenamefont {Gohda},\ and\ \citenamefont
  {Tsuneyuki}}]{PhysRevB.89.094515}%
  \BibitemOpen
  \bibfield  {author} {\bibinfo {author} {\bibfnamefont {M.}~\bibnamefont
  {Kawamura}}, \bibinfo {author} {\bibfnamefont {Y.}~\bibnamefont {Gohda}}, \
  and\ \bibinfo {author} {\bibfnamefont {S.}~\bibnamefont {Tsuneyuki}},\ }\href
  {\doibase 10.1103/PhysRevB.89.094515} {\bibfield  {journal} {\bibinfo
  {journal} {Phys. Rev. B}\ }\textbf {\bibinfo {volume} {89}},\ \bibinfo
  {pages} {094515} (\bibinfo {year} {2014})}\BibitemShut {NoStop}%
\end{thebibliography}%

\vspace{1em}
\noindent
\textbf{Acknowledgments:}
Work at the University of Florida performed under the auspices of U.S. Department of Energy Basic Energy Sciences under Contract No.\ DE-SC-0020385.
We thank S. Tkachev (GSECARS, University of Chicago) for sample gas loading for the x-ray diffraction measurements.
We acknowledge Curtis Kenny Benson, HPCAT, for technical assistance.
A.C.H.\ and R.G.H.\ acknowledge support from the National Science Foundation under award PHY-1549132 (Center for Bright Beams) and award DMR-2118718.
Y.K.V.\ acknowledges the support from DOE-NNSA award DE-NA0003916.
R.S.K.\ and R.J.H.\ acknowledge support from the U.S.\ National Science Foundation (DMR-1933622 and DMR-2119308).
X-ray diffraction measurements were performed at HPCAT (Sector 16), Advanced Photon Source (APS), Argonne National Laboratory. HPCAT operations are supported by the DOE-National Nuclear Security Administration (NNSA) Office of Experimental Sciences.
The beamtime was made possible by the Chicago/DOE Alliance Center (CDAC), which is supported by DOE-NNSA (DE-NA0003975).
Use of the gas loading system was supported by COMPRES under NSF Cooperative Agreement EAR-1606856 and by GSECARS through NSF grant EAR-1634415 and DOE grant DE-FG02-94ER14466.
The Advanced Photon Source is a DOE Office of Science User Facility operated for the DOE Office of Science by Argonne National Laboratory under Contract No. DE-AC02-06CH11357.
We thank M.\ Trenary for useful discussions and comments on the manuscript.
High pressure equipment development at the University of Florida was supported by National Science Foundation CAREER award DMR-1453752.

\vspace{1em}
\noindent
\textbf{Author Contributions Statement:}
The initial computational $T_c$ estimates that motivated this study were performed by Y.\ Quan.
High pressure electrical resistivity measurements were performed by J.\ Lim with the assistance of S.\ Sinha.
DFT calculations and analysis were performed by A.\ C.\ Hire with contributions from S.\ R.\ Xie at an early stage.
Samples were synthesized and characterized at ambient pressure by J.\ S.\ Kim and G.\ R.\ Stewart.
High pressure x-ray measurements and analyses were performed by J.\ Lim with the assistance of R.\ S.\ Kumar, D.\ Popov, and C.\ Park, and R.\ J.\ Hemley.
X-ray data analysis was performed by J.\ Lim and A.\ C.\ Hire.
Y.\ K.\ Vohra lead the fabrication of designer diamond anvils that were used for a portion of the high pressure electrical resistivity measurements.
The paper was written primarily by J.\ Lim, A.\ Hire, J.\ J.\ Hamlin, R.\ G.\ Hennig, P.\ J.\ Hirschfeld, and G.\ R.\ Stewart, with contributions and comments from the other authors.
The study was conceived and organized by J.\ J.\ Hamlin, R.\ G.\ Hennig, P.\ J.\ Hirschfeld, and G.\ R.\ Stewart.

\vspace{1em}
\noindent
\textbf{Competing Interests Statement:}
The authors declare no competing interests.
\end{document}


\title{Supplemental material: Creating superconductivity in \ch{WB2} through pressure-induced\\ metastable planar defects}

\author{J.\ Lim}
\thanks{These two authors contributed equally to this work.}
\affiliation{Department of Physics, University of Florida, Gainesville, Florida 32611, USA}
\author{A.\ C.\ Hire}
\thanks{These two authors contributed equally to this work.}
\affiliation{Department of Materials Science and  Engineering, University of Florida, Gainesville, Florida 32611, USA}
\affiliation{Quantum Theory Project, University of Florida, Gainesville, Florida 32611, USA}
\author{Y.\ Quan}
\affiliation{Department of Physics, University of Florida, Gainesville, Florida 32611, USA}
\affiliation{Department of Materials Science and  Engineering, University of Florida, Gainesville, Florida 32611, USA}
\affiliation{Quantum Theory Project, University of Florida, Gainesville, Florida 32611, USA}
\author{J.\ S.\ Kim}
\affiliation{Department of Physics, University of Florida, Gainesville, Florida 32611, USA}
\author{S.\ R.\ Xie}
\affiliation{Department of Materials Science and  Engineering, University of Florida, Gainesville, Florida 32611, USA}
\affiliation{Quantum Theory Project, University of Florida, Gainesville, Florida 32611, USA}
\author{S.\ Sinha}
\affiliation{Department of Physics, University of Florida, Gainesville, Florida 32611, USA}
\author{R.\ S.\ Kumar}
\affiliation{Department of Physcis, Chemistry, and Earth and Environmental Sciences, University of Illinois Chicago, Chicago, Illinois 60607, USA}
\author{D.\ Popov}
\affiliation{HPCAT, X-ray Science Division, Argonne National Laboratory, Argonne, Illinois 60439, USA}
\author{C.\ Park}
\affiliation{HPCAT, X-ray Science Division, Argonne National Laboratory, Argonne, Illinois 60439, USA}
\author{R.~J.~Hemley}
\affiliation{Department of Physcis, Chemistry, and Earth and Environmental Sciences, University of Illinois Chicago, Chicago, Illinois 60607, USA}
\author{Y.\ K.\ Vohra}
\affiliation{Department of Physics, University of Alabama at Birmingham, Birmingham, Alabama 35294, USA}
\author{J.\ J.\ Hamlin}
\email{jhamlin@ufl.edu}
\affiliation{Department of Physics, University of Florida, Gainesville, Florida 32611, USA}
\author{R.\ G.\ Hennig}
\affiliation{Department of Materials Science and  Engineering, University of Florida, Gainesville, Florida 32611, USA}
\affiliation{Quantum Theory Project, University of Florida, Gainesville, Florida 32611, USA}
\author{P.\ J.\ Hirschfeld}
\affiliation{Department of Physics, University of Florida, Gainesville, Florida 32611, USA}
\author{G.\ R.\ Stewart}
\affiliation{Department of Physics, University of Florida, Gainesville, Florida 32611, USA}

\maketitle
\section{XRD of WB$_2$ hR6 phase}
\begin{figure}[h!]
  \includegraphics[width=\linewidth]{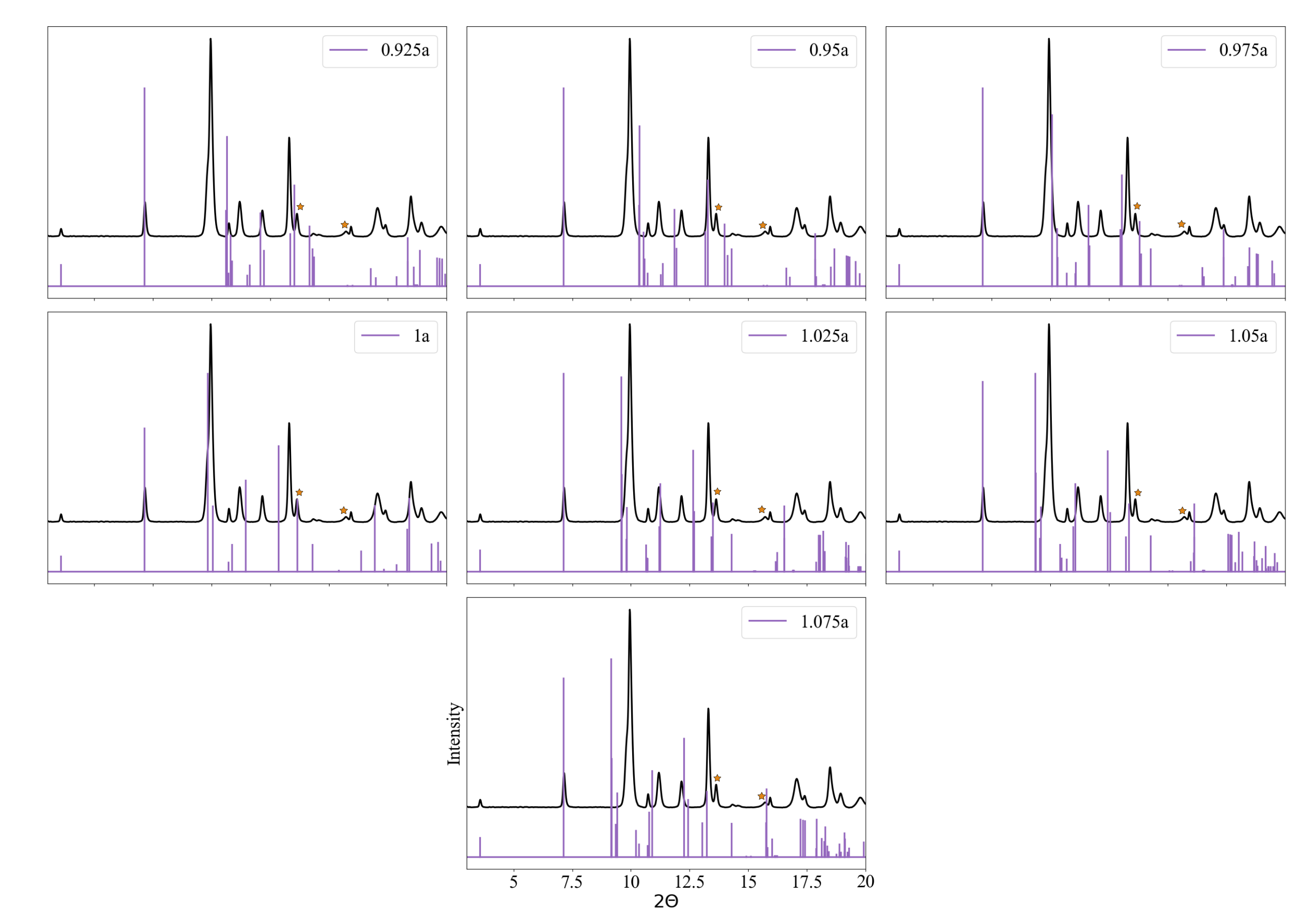}
  \caption{Experimentally measured XRD pattern of WB$_2$ and theoretically calculated peaks of the strained hR6 phase at 85~GPa. The $a$ lattice vector of the structure was strained whereas the $c$ lattice vector was kept unchanged. The legend in each plot indicates the amount by which the $a$ lattice parameter was changed with respect to the DFT relaxed structure.}
  \label{WB2_hR6_strained_85}
\end{figure}

\begin{figure}[h!]
  \includegraphics[width=\linewidth]{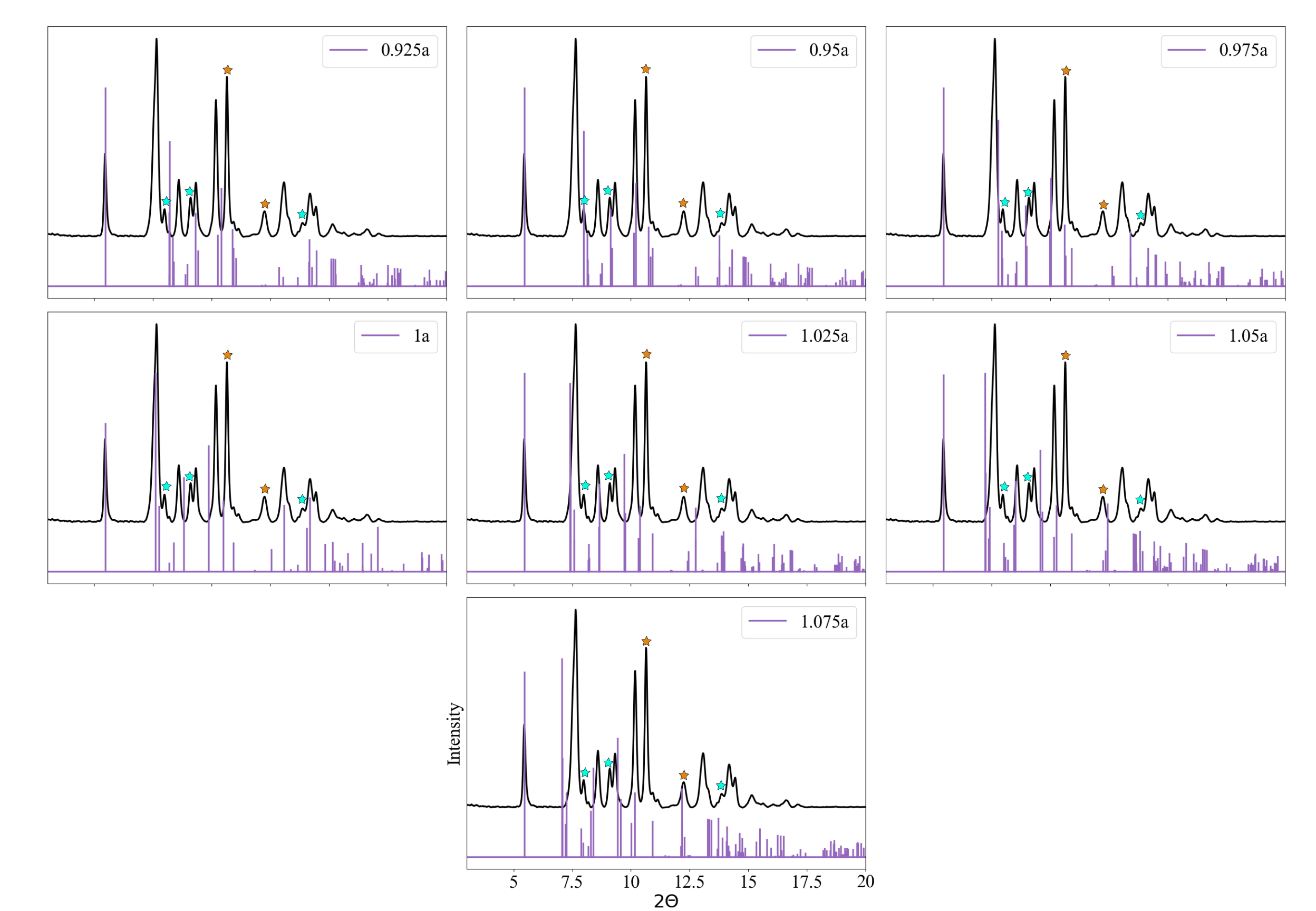}
  \caption{Experimentally measured XRD pattern of WB$_2$ and theoretically calculated peaks of the strained hR6 phase at 145~GPa. The \lq a\rq~ lattice vector of the structure was strained whereas the \lq c\rq~lattice vector was kept unchanged. The legend in each plot indicates the amount by which the \lq a\rq~lattice parameter was changed with respect to the DFT relaxed structure.}
  \label{WB2_hR6_strained_145}
\end{figure}

The XRD peaks of the WB$_2$ hP12 and hR6 phases are quite similar, such that the broadening of the experimentally measured XRD peaks at high pressures leads to difficulty in conclusively identifying the stable phase.
To check whether a strained lattice of the hR6 phase can account for the experimentally observed high-pressure XRD pattern, we perform the following analysis. We strain the DFT relaxed lattice parameters of the hR6 phase and compare the resulting theoretical XRD pattern against the experimental data. The $c$ lattice parameter was kept unchanged from its DFT value because the experimental XRD peaks that contain contributions from the planes perpendicular to the $c$-axis match almost perfectly with the theoretical peaks. We performed this analysis at two pressures, \SI{85}{GPa} and \SI{145}{GPa}, as shown in Figs.\ref{WB2_hR6_strained_85} and \ref{WB2_hR6_strained_145}. In these figures, the peaks denoted by the star symbol are due to the Re-gasket (cyan) and the Ne pressure-transmitting medium (orange). We observe that the strained hR6 lattice fails to produce the experimentally observed pattern at \SI{85}{GPa} and \SI{145}{GPa}. This failure of the strained lattice reveals the absence of significant amounts of the hR6 phase. We further rule out the presence of the hR6 phase with the help of the theoretically calculated critical temperatures. According to our calculations, the hR6 phase has a critical temperature of less than \SI{5}{K} up to \SI{100}{GPa}, which is far less than the observed $T_\mathrm{c}$ of \SI{17}{K} at \SI{90}{GPa}.
\clearpage
\newpage
\section{WB$_2$ Stacking fault and Twin boundary}
\begin{figure}[h!]
  \includegraphics[width=0.95\linewidth]{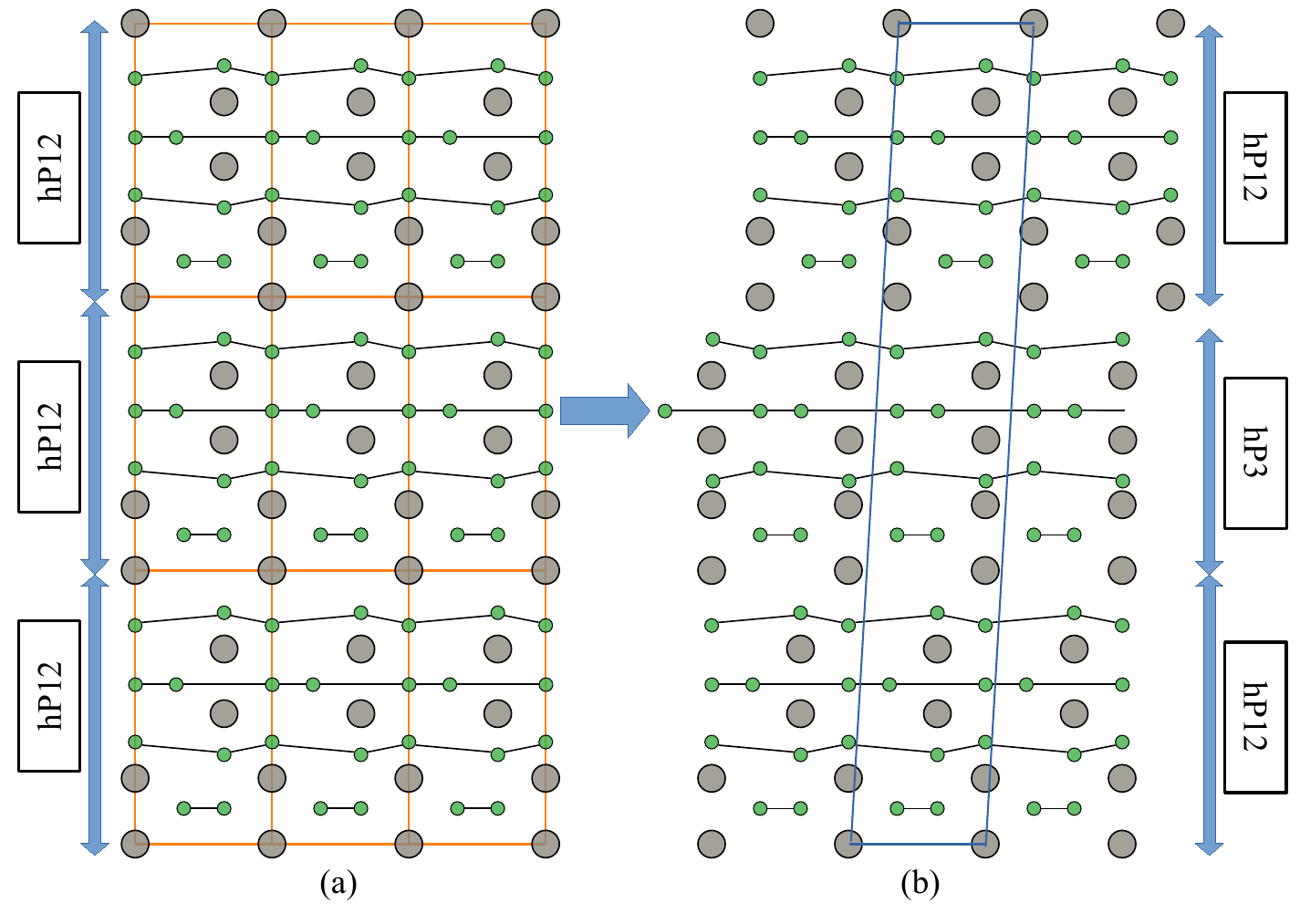}
  \caption{Formation of stacking fault by sliding of planes. We expect that the buckled Boron layers in the hP3 region will unbuckle and shift by 1/3 of the \lq a\rq~ lattice vector to lower the energy of the defect structure.}
  \label{WB2_SF}
\end{figure}

\begin{figure}[h!]
  \includegraphics[width=0.95\linewidth]{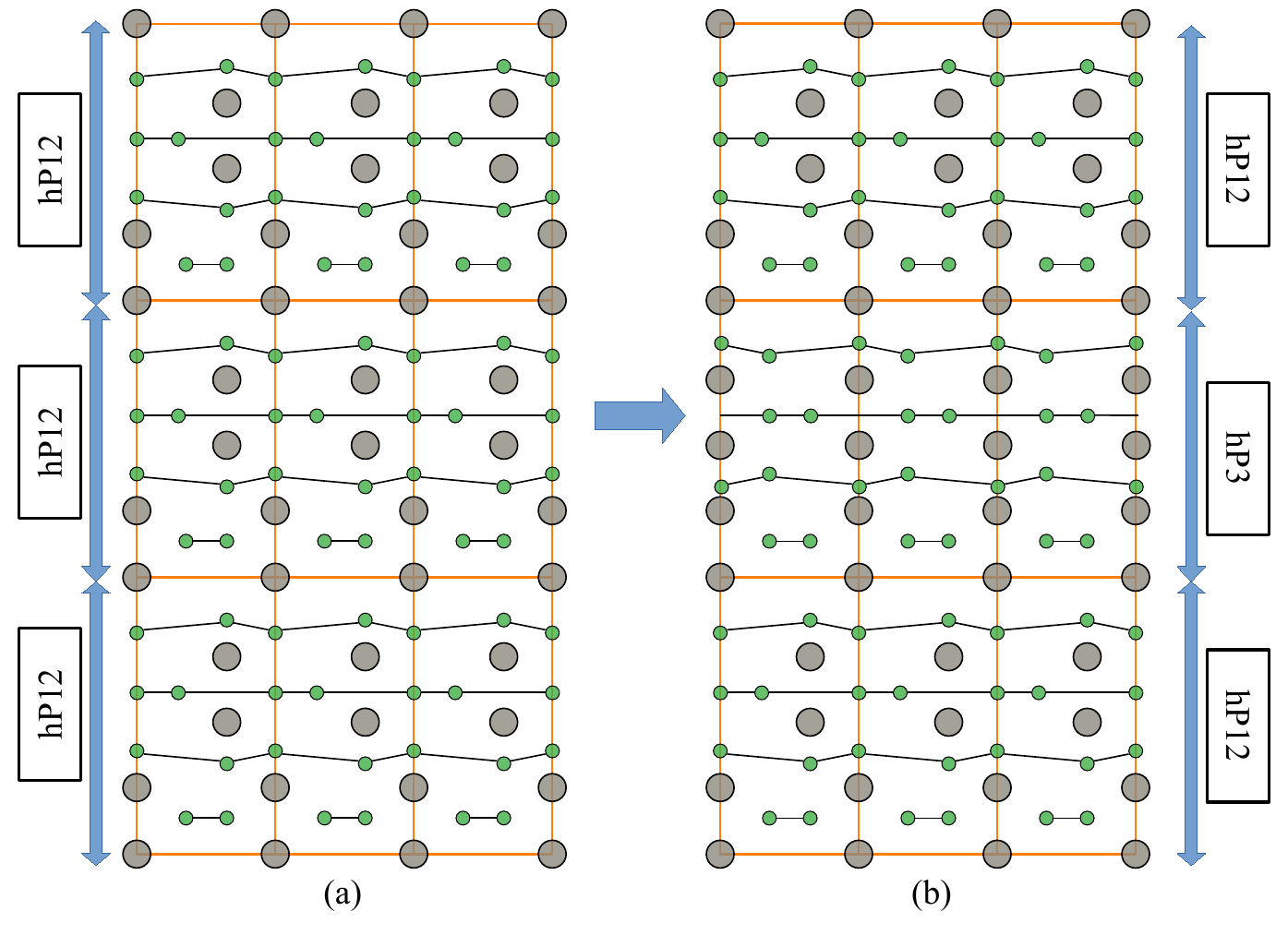}
  \caption{Formation of twin boundary by sliding of planes. We expect that the buckled Boron layers in the hP3 region will unbuckle and shift by 1/3 of the \lq a\rq~lattice vector to lower the energy of the defect structure.}
  \label{WB2_TB}
\end{figure}

As the pressure applied to the WB$_2$ samples increases, stacking faults and twin boundary defects can form inside the sample because of mechanical deformation. Here, with the help of Figures \ref{WB2_SF} and \ref{WB2_TB}, we present how simulation cells containing these planar defects can be created in the WB$_2$ hP12 phase by sliding W and B planes. Figures \ref{WB2_SF}(a) and \ref{WB2_TB}(a) show the perfect hP12 unit cells. A cell with a stacking fault is created by sliding the appropriate W and B planes along the $[1/3, 2/3, 0]$ vector in the hP12 cell, as shown in Figure \ref{WB2_SF}(b). In Figure \ref{WB2_TB}(b), the simulation cell of the twin boundary is formed by sliding only two successive W and B planes along the $[1/3, 2/3, 0]$ vector. For calculating the defect formation energy depicted in Figure 4(i), we use simulation cells equivalent to four hP12 unit cells.

\clearpage

\section{Electronic structure of WB$_2$ under pressure}
\begin{figure}[!ht]
  \includegraphics[width=\columnwidth]{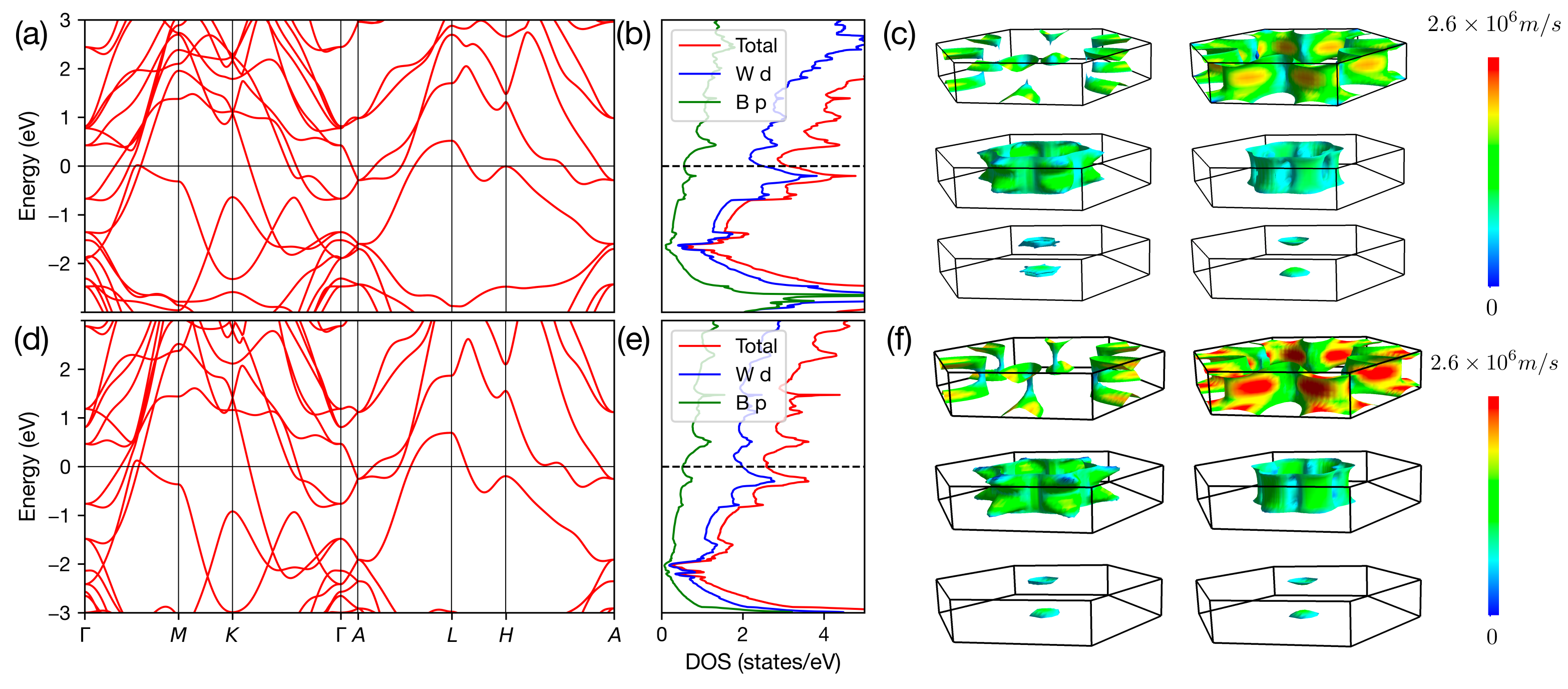}
  \caption{Band structures, density of states and Fermi surfaces of WB$_2$ in P6$_3$/mmc structure under 0 and 100 GPa. (a) to (c) correspond to WB$_2$ under 0 GPa and (d) to (f) correspond to WB$_2$ under 100 GPa.}
  \label{fig:WB2es}
\end{figure}

The electronic structures of WB$_2$ in the experimental space group P6$_3$/mmc under 0 and 100 GPa are shown in Fig.~\ref{fig:WB2es}. The states near the Fermi level
are mostly tungsten $d$ states and they are
important in understanding the superconductivity of  WB$_2$. Pressure usually has the effect of broadening
bandwidth. However,
the broadening is not uniform. For example, the band that touches the Fermi level at $H$ point under 0 GPa moves to 0.2 eV below the Fermi level under 100 GPa, while the two bands that merge at $A$ point remain 0.3 eV below the Fermi level. Overall, the density of
states evolve smoothly with pressure and the Fermi level falls inside a DOS valley, contrary
to the sudden emergence of superconductivity at 50 GPa
observed in experiments. In addition, the
DFT-estimated electron-phonon coupling of WB$_2$
in the P6$_3$/mmc phase is small and cannot
account for the experimental T$_c$, 
which suggests that the superconductivity might
not be due to the P6$_3$/mmc phase. On the other hand,
the electron-phonon coupling of WB$_2$ in the MgB$_2$
structure is strong, pointing to possible
existence of local MgB$_2$-like structures in certain regions of the experimental sample that is responsible for the superconductivity of WB$_2$ under pressure.

\section{Pressure dependence of phonon dispersion of WB$_2$ in the hP$_3$ phase}
\begin{figure}[!ht]
\includegraphics[width=\columnwidth]{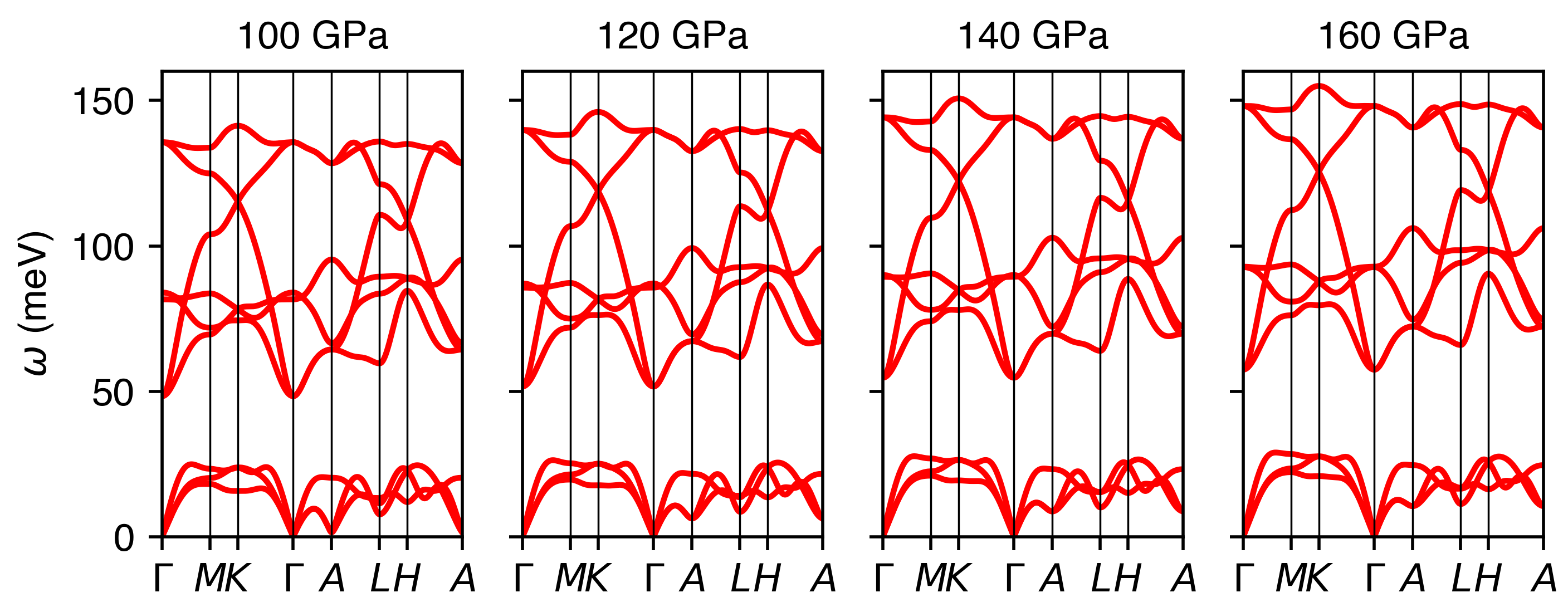}
\caption{Pressure dependence of the phonon dispersion of WB$_2$ in the hP$_3$ phase. }
\label{phonon}
\end{figure}

In recent years, high pressure experiments have played
an important role in expanding the frontier of phases
of materials that can be stabilized. Structures that
are otherwise chemically or dynamically unstable under
the ambient condition can be easily pressurized into
existence. In this section, we present the pressure
dependence of the phonon dispersion of WB$_2$, see~\ref{phonon}. The
phonon spectrum is calculated by running a QE
calculation on a coarse mesh first and then an
EPW calculation is carried out to interpolated 
from the coarse k- and q-meshes to much
finer meshes. We have assumed harmonic approximation
which can already provide important guidance for
understanding the experimental data. Anharmonic
effects might have some impact on the exact
pressure at which the hP$_3$ phase becomes unstable. However,
the issue of anharmonicity is out of the scope of our
current manuscript. 

\clearpage
\newpage

\section{Temperature-dependent electrical resistivity of \ch{WB2} under decompression in Run 2}
\begin{figure}[!ht]
  \includegraphics[width=0.7\columnwidth]{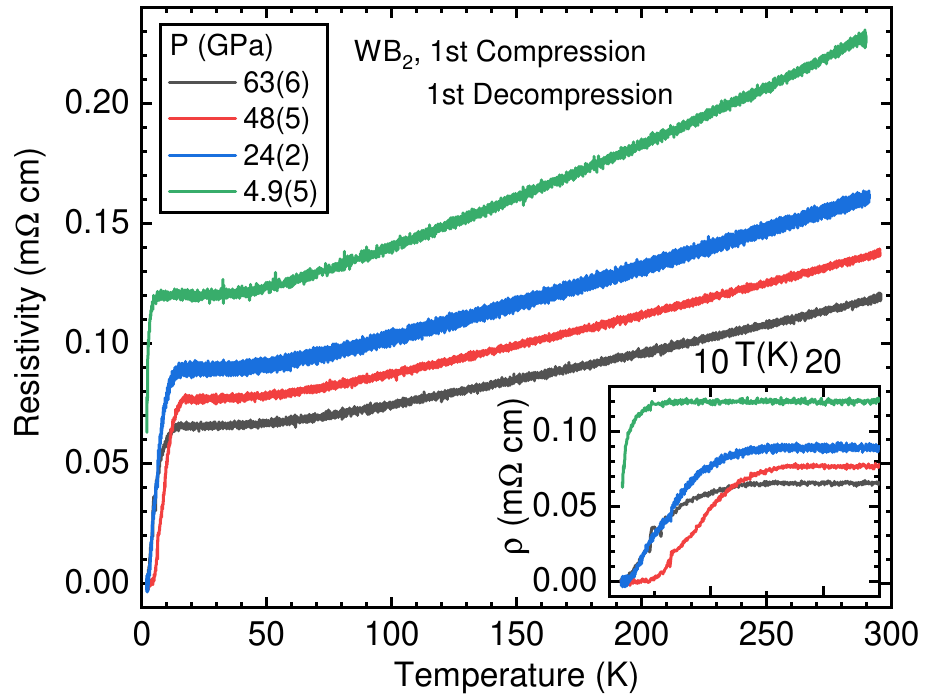}
  \caption{Temperature-dependent electrical resistivity curves of \ch{WB2} during decompression from 63 down to \SI{4.9}{GPa} in Run 2.}
  \label{WB2_unloading}
\end{figure}

Figure~\ref{WB2_unloading} shows the decompression behavior of temperature-dependent electrical resistivity of \ch{WB2} in Run 2. The superconducting temperature $T_{c}$ (90\%) at \SI{63}{GPa} (black curve, first compression) is $\sim$\SI{10}{K} in consistent with Run 1 as shown in Fig.~1d. When lowering the pressure down to \SI{48}{GPa} (red curve), interestingly $T_{c}$ rather increases to $\sim$\SI{13}{K} and starts to decrease to $\sim$\SI{3}{K} at further lowering to \SI{4.9}{GPa}, which shows the irreversible behavior of $T_{c}(P)$ under decompression.
This shows that the superconducting phase of \ch{WB2} is metastable at low pressure.

\clearpage
\newpage

\section{Temperature-dependent electrical resistivity of \ch{WB2} under the second compression in Run 2}
\begin{figure}[!ht]
  \includegraphics[width=0.7\columnwidth]{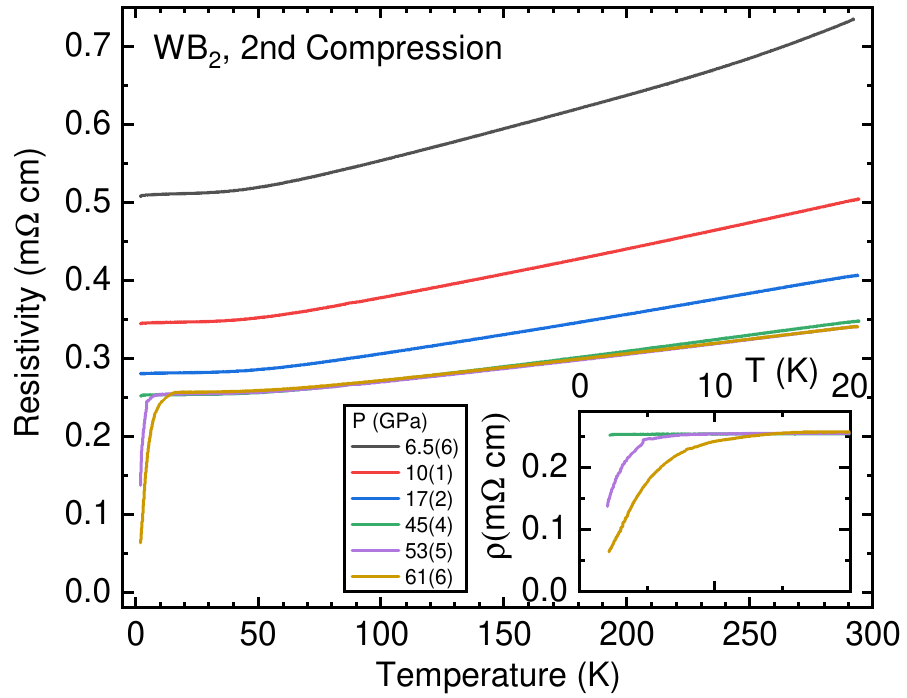}
  \caption{Temperature-dependent electrical resistivity curves of \ch{WB2} during the second compression from 6.5 up to \SI{61}{GPa} in Run 2.}
  \label{WB2_2ndComp}
\end{figure}

Figure~\ref{WB2_2ndComp} shows the second compression behavior of temperature-dependent electrical resistivity of \ch{WB2} after lowering to ambient pressure (opening the DAC) from the first decompression (Fig. S7) in Run 2. The superconductivity begins to appear from \SI{53}{GPa} (purple curve) with $T_c \sim \SI{4}{K}$ in consistent with Run 1 as shown in Fig.~1d. The results indicate that the planar defects or twin boundaries responsible for the superconductivity is only (meta)stable under high pressure.

\clearpage
\newpage

\section{High-pressure XRD patterns of \ch{WB2} under nonhydrostatic pressure condition}
\begin{figure}[!ht]
  \includegraphics[width=\columnwidth]{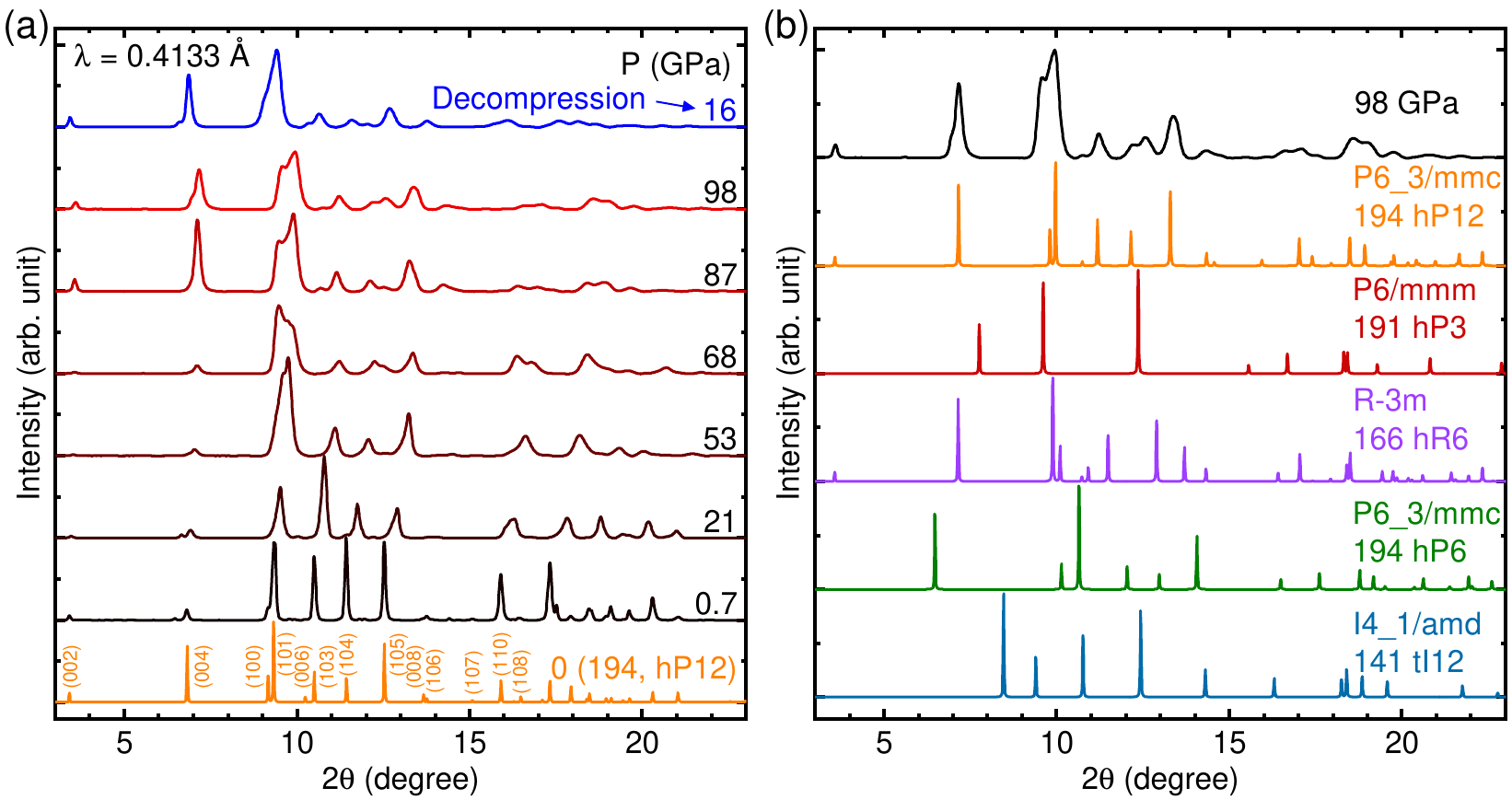}
  \caption{(a) High pressure XRD patterns of \ch{WB2} to \SI{98}{GPa} under nonhydrostatic condition (no pressure medium) (b) Comparison of the XRD pattern at \SI{98}{GPa} to those of five competing phases.}
  \label{WB2_XRDnonhydrostatic}
\end{figure}

Figure~\ref{WB2_XRDnonhydrostatic} shows high-pressure XRD measurements under nonhydrostatic pressure condition to \SI{98}{GPa}.
Without using any pressure medium, the pressure chamber is filled with the \ch{WB2} samples in these measurements.
The XRD patterns in Fig.~\ref{WB2_XRDnonhydrostatic}(a) match with hP12 structure (orange line, calculated at \SI{0}{GPa}) except for few minor additional peaks appearing above \SI{50}{GPa} that could be assigned to either hP3 or hR6 phase as possible local defects.
However, other large peaks associated with the hP3 and hR6 structures (e.g.\ peaks near 7-8 degrees) are entirely missing from the experimental data.
Figure~\ref{WB2_XRDnonhydrostatic}(b) shows the XRD pattern at \SI{98}{GPa} compared to those of the five competing phases in in Fig.~3 of the main text.
The data are not not of sufficient quality to permit a full quantitative Rietveld refinement,
but nonetheless, the results confirm that the bulk \ch{WB2} sample still remains predominantly in hP12 phase up to \SI{98}{GPa} under non-hydrostatic conditions.
The results are consistent with the quasi-hydrostatic XRD measurements where \ch{Ne} was used as the pressure medium (Fig.~2, main text).

